\documentclass[onecolumn,superscriptaddress,secnumarabic,amssymb,amsmath,nobibnotes,aps,prd,nofootinbib,altaffilletter]{revtex4}
\usepackage{amsmath}
\usepackage{amsfonts,color}
\usepackage{amssymb,float}
\usepackage[utf8]{inputenc}
\usepackage{color}
\usepackage{enumitem}
\usepackage{graphicx}
\usepackage{appendix}
\usepackage{hyperref}
\usepackage{accents}
\usepackage{lipsum}
\usepackage{xcolor}
\usepackage{orcidlink}
\usepackage{epsf}
\usepackage{bm}
\usepackage{epstopdf}
\usepackage{natbib}
\usepackage{rotating}
\usepackage{pdflscape}
\usepackage{adjustbox}
\usepackage{color}
\usepackage{dcolumn}
\usepackage{ulem}

\newcommand{\udt}[3]{#1^{#2}_{\phantom{#2}#3}}

\newcommand{\lc}[1]{\accentset{\circ}{#1}}

\begin{document}

\title{Stability analysis for cosmological models in $f(T,B)$ gravity}

\author{Geovanny A. Rave Franco}
\email{geovanny.rave@ciencias.unam.mx}
\affiliation{Instituto de Ciencias Nucleares, Universidad Nacional Aut\'onoma de M\'exico, Circuito Exterior C.U., A.P. 70-543, M\'exico D.F. 04510, M\'exico.}

\author{Celia Escamilla-Rivera\orcidlink{0000-0002-8929-250X}}
\email{celia.escamilla@nucleares.unam.mx}
\affiliation{Instituto de Ciencias Nucleares, Universidad Nacional Aut\'onoma de M\'exico, Circuito Exterior C.U., A.P. 70-543, M\'exico D.F. 04510, M\'exico.}
	
\author{Jackson Levi Said\orcidlink{0000-0002-8929-250X}}
\email{jackson.said@um.edu.mt}
\affiliation{Institute of Space Sciences and Astronomy, University of Malta, Msida, MSD 2080, Malta}
\affiliation{Department of Physics, University of Malta, Msida, MSD 2080, Malta}


\begin{abstract}

In this paper we study cosmological solutions of the $f(T,B)$ gravity using dynamical system analyses. For this purpose, we consider cosmological viable functions of $f(T,B)$ that are capable of reproducing the dynamics of the Universe. We present three specific models of $f(T,B)$ gravity which have a general form of their respective solutions by writing the equations of motion as an autonomous system. Finally, we study its hyperbolic critical points and general trajectories in the phase space of the resulting dynamical variables which turn out to be compatible with the current late-time observations.

\end{abstract}

\maketitle

\section{Introduction}

The $\Lambda$CDM cosmological model is supported by overwhelming observational evidence in describing the evolution of the Universe at all cosmological scales \cite{misner1973gravitation,Clifton:2011jh} which is achieved by the inclusion of matter beyond the standard model of particle physics. This takes the form of dark matter which acts as a stabilizing agent for galactic structures \cite{Baudis:2016qwx,Bertone:2004pz} and materializes as cold dark matter particles, while dark energy is represented by the cosmological constant \cite{Peebles:2002gy,Copeland:2006wr} and produces the measured late-time accelerated expansion \cite{Riess:1998cb,Perlmutter:1998np}. However, despite great efforts, internal consistency problems persist with the cosmological constant \cite{RevModPhys.61.1}, as well as a severe lack of direct observations of dark matter particles \cite{Gaitskell:2004gd}.

On the other hand, the effectiveness of the $\Lambda$CDM model has also become an open problem in recent years. At its core, the $\Lambda$CDM model was convinced to describe Hubble data but the so-called $H_0$ tension problem calls this into question where the observational discrepancy between model independent measurements \cite{Riess:2019cxk,Wong:2019kwg} and predicted \cite{Aghanim:2018eyx,Ade:2015xua} values of $H_0$ from the early-Universe appears to be growing \cite{Gomez-Valent:2019lny}. While measurements from the tip of the red giant branch (TRGB, Carnegie-Chicago Hubble Program) point to a lower $H_0$ tension, the issue may ultimately be resolved by future observations which may involve more exotic measuring techniques such as the use of gravitational wave astronomy \cite{Graef:2018fzu,Abbott:2017xzu} with the LISA mission \cite{Baker:2019nia,2017arXiv170200786A}. However, it may also be the case that physics beyond general relativity (GR) are at play here.

The use of autonomous differential equations to investigate the cosmological dynamics of modified theories of gravity has been shown to be a powerful tool in elucidating cosmic evolution within these possible models of gravity \cite{Bahamonde:2017ize}. These analyses can reveal the underlying stability conditions of a theory from which it may be possible to constrain possible models on theoretical grounds alone. Theories beyond GR come in many different flavours \cite{Clifton:2011jh,Capozziello:2011et} where many are designed to impact the currently observed late-time cosmology dynamics. By and large, the main trust of these theories comes in the form of extended theories of gravity \cite{Sotiriou:2008rp,Faraoni:2008mf,Capozziello:2011et} which build on GR with correction factors that dominate for different phenomena. However, these are collectively all based on a common mechanism by which gravitation is expressed through the Levi-Civita connection, i.e. that gravity is communicated by means of the curvature of spacetime \cite{misner1973gravitation}. In fact, it is the geometric connection which expresses gravity while the metric tensor quantifies the amount of deformation present \cite{nakahara2003geometry}. This is not the only choice where torsion has become an increasingly popular replacement and produced a number of well-motivated theories \cite{Aldrovandi:2013wha,Cai:2015emx,Krssak:2018ywd}.

Teleparallel Gravity (TG) collectively embodies the class of theories of gravity in which gravity is expressed through torsion through the teleparallel connection \cite{Weitzenbock1923}. This connection is torsion-ful while being curvatureless and satisfying the metricity condition. Naturally, all curvature quantities calculated with this connection will vanish irrespective of metric components. Indeed, the Einstein-Hilbert which is based on the Ricci scalar $\lc{R}$ (over-circles represent quantities calculated with the Levi-Civita connection) vanishes when calculated with the teleparallel connection, i.e. in general $R=0$ while $\lc{R}\neq 0$. Moreover, the identical dynamical equations can be arrived at in TG by replacing the Einstein-Hilbert action with it's so-called torsion scalar $T$. By making this substitution, we produce the \textit{Teleparallel equivalent of General Relativity} (TEGR), which differs from GR by a boundary term $B$ in its Lagrangian.

The boundary term in TEGR consolidates the fourth-order contributions to many beyond GR theories. By extracting these contributions into a separate scalar, TEGR will have a meaningful and novel impact on extended theories and produce an impactful difference in the predictions of such theories. The most direct result of this fact will be that TG will produce a much broader plethora of theories in which dynamical equations are second order. This is totally different to the severely limited Lovelock theorem in curvature based theories \cite{Lovelock:1971yv}. In fact, TG can produce a large landscape in which second-order field equations are produced \cite{Gonzalez:2015sha,Bahamonde:2019shr}. TG also has a number of other attractive features such as its likeness to Yang-mills theory \cite{Aldrovandi:2013wha} offering a particle physics perspective to the theory, the possibility of it giving a definition to the gravitational energy-momentum tensor\cite{Blixt:2018znp,Blixt:2019mkt}, and that it does not necessitate the introduction of a Gibbons--Hawking--York boundary term to produce a well-defined Hamiltonian structure, among others.

Taking the same reasoning as in $f(\lc{R})$ gravity \cite{Sotiriou:2008rp,Faraoni:2008mf,Capozziello:2011et}, TEGR can be straightforwardly generalized to produce $f(T)$ gravity \cite{Ferraro:2006jd,Ferraro:2008ey,Bengochea:2008gz,Linder:2010py,Chen:2010va,Bahamonde:2019zea}. $f(T)$ gravity is generally second order due to the weakened Lovelock theorem in TG and has shown promise in several key observational tests \cite{Cai:2015emx,Nesseris:2013jea,Farrugia:2016qqe,Finch:2018gkh,Farrugia:2016xcw,Iorio:2012cm,Ruggiero:2015oka,Deng:2018ncg}. However to fully embrace the TG generalization of $f(\lc{R})$ gravity, we must consider the $f(T,B)$ generalization of TEGR \cite{Bahamonde:2015zma,Capozziello:2018qcp,Bahamonde:2016grb,Farrugia:2018gyz,Bahamonde:2016cul,Bahamonde:2016cul,Wright:2016ayu}. In this scenario the second and fourth order contributions to $f(\lc{R})$ gravity are decoupled while this subclass becomes a particular limit of the arguments $T$ and $B$, namely $f(\lc{R})=f(-T+B)$. $f(T,B)$ gravity is an interesting theory of gravity and has shown promise in terms of solar system tests and the weak field regime \cite{Farrugia:2020fcu,Capozziello:2019msc,Farrugia:2018gyz}, as well as cosmologically both in terms of its theoretical structure \cite{Bahamonde:2015zma,Bahamonde:2016grb,Bahamonde:2016cul,Bahamonde:2015zma} and confrontation with observational data \cite{Escamilla-Rivera:2019ulu}.

In this work, we explore the structure of $f(T,B)$ gravity through the dynamical systems approach in the cosmological context of a homogeneous and isotropic Universe using the Friedmann–Lema\^{i}tre–Robertson–Walker metric (FLRW). This kind of system has been used to study higher-order modified teleparallel gravity that add a scalar field $\phi$ depending on the boundary term \cite{Bahamonde:2016grb}, where the stability conditions for a number of exact solutions for an FLRW background are studied for a number of solutions such as de Sitter Universe and ideal gas solutions. In Ref.\cite{Karpathopoulos:2017arc}, a number of important reconstructions were presented together with their dynamical system evolution. This work is also interesting because they compare some of their results with Supernova type 1a data using a chi-square approach. Another interesting approach to determining and studying solutions is that of using Noether symmetries Ref.\cite{Capozziello:2014bna}, which have \cite{Bahamonde:2016grb} shown great promise in producing new cosmological solutions that admit more desirable cosmologies. On the other hand, in Ref.\cite{Bahamonde:2016cul} the all important cosmological thermodynamics has been explored as well as the mater perturbations, which was complemented by background reconstructions of further cosmological solutions giving a rich literature of models together with Ref.\cite{Pourbagher:2019zhq}. Ref.\cite{Zubair:2018wyy} then developed the $f(T,B)$ cosmology energy conditions which can give important information about regions of validity of the models.

In the present study, the cosmic acceleration dynamics is reproduced only by a non-canonical $\phi$ that mimics the $\Lambda$ term. In our case, we introduce a $f(T,B)$ dark energy which is fluid-like in order to obtain a richer population of stability points that can be constrained by current observational surveys. We do this by first introducing briefly the technical details of $f(T,B)$ gravity in section~\ref{sec: cosmology_fTB} and then discussing its dynamical treatment in section~\ref{f_T_B_dynamic_struc}. In section~\ref{stabil_method} we lay out the methodology of the analysis which includes the methods by which the analysis is conducted. The $f(T,B)$ gravity dynamical analysis is then realized in section~\ref{dynami_ana} where the core results for each of the models is presented. Finally, we close in section~\ref{conc} with a summary of our conclusions. In all that follows, Latin indices are used to refer to Minkowski space coordinates, while Greek indices refer to general manifold coordinates.

\section{\texorpdfstring{$f(T,B)$}{f(T,B)} cosmology}
\label{sec: cosmology_fTB}

We start by considering a flat homogeneous and isotropic FLRW metric in Cartesian coordinates with an absorbed lapse function ($N=1$) as (e.g through Ref.\cite{misner1973gravitation})
\begin{equation}\label{metric}
{\rm d}s^2=-{\rm d}t^2+a(t)^2({\rm d}x^2+{\rm d}y^2+{\rm d}z^2)\,,
\end{equation}
where $a(t)$ is the scale factor. Also, we choose an arbitrary mapping over $\tilde{f}(T,B) \rightarrow -T + f(T,B)$, which obeys the diffeomorphism invariance. As shown in \cite{Escamilla-Rivera:2019ulu}, the choice of Lagrangian where $\tilde{f}(T,B) \rightarrow -T + f(T,B)$ represents an arbitrary Lagrangian over the torsion scalar and boundary term is diffomorphism invariant. In this proposal our choice of tetrad is given by
\begin{equation}\label{FLRW_tetrad}
\udt{e}{a}{\mu}=\mbox{diag}(1,a(t),a(t),a(t))\,,
\end{equation}
which reproduces the metric in Eq.(\ref{metric}) and observes the symmetries of TG. In this spacetime, the torsion scalar can be given explicitly as
\begin{equation}\label{torsionscalar_frw}
T = 6H^2\,,
\end{equation}
while the boundary term is given by
\begin{equation}\label{boundaryscalar_frw}
B = 6\left(3H^2+\dot{H}\right)\,,
\end{equation}
which combine to produce the well known Ricci scalar of the flat FLRW metric $\lc{R}=-T+B=6\left(\dot{H}+2H^2\right)$ (where again over-circles again represent quantities determined with the Levi-Civita connection).

After the above considerations over the geometry, our field equations for a universe filled with a perfect fluid are 
\begin{eqnarray}
&&-3H^2\left(3f_B + 2f_T\right) + 3H\dot{f}_B - 3\dot{H} f_B + \frac{1}{2}f = \kappa^2\rho\label{Friedmann_1}\,, \\
&&-\left(3H^2+\dot{H}\right)\left(3f_B + 2f_T\right) - 2H\dot{f}_T + \ddot{f}_B + \frac{1}{2}f = -\kappa^2 p\label{Friedmann_2}\,,
\end{eqnarray}
where $\rho$ and $p$ represent the energy density and pressure of a perfect fluid whose equation of state is $p= \omega \rho$, respectively. These modified Friedmann equations show explicitly how a linear boundary contribution to the Lagrangian would act as a boundary term while other contributions of $B$ would contribute nontrivially to the dynamics of these equations. 

We can rewrite Eqs.(\ref{Friedmann_1},\ref{Friedmann_2}) by considering the modified TEGR components contained in the effective fluid contributions
\begin{eqnarray}
3H^2 &=& \kappa^2 \left(\rho +\rho_{\text{eff}}\right)\,,\\
3H^2+2 \dot{H} &=& -\kappa^2\left(p+p_{\text{eff}}\right)\,,
\end{eqnarray}
where 
\begin{eqnarray}
&&\kappa^2 \rho_{\text{eff}} = 3H^2\left(3f_B + 2f_T\right) - 3H\dot{f}_B + 3\dot{H}f_B - \frac{1}{2}f\,, \label{eq:friedmann_mod}\\
&&\kappa^2 p_{\text{eff}} = \frac{1}{2}f-\left(3H^2+\dot{H}\right)\left(3f_B + 2f_T\right)
-2H\dot{f}_T+\ddot{f}_B\,.
\end{eqnarray}
The latter equation can be combined to obtain
\begin{equation}
2\dot{H}=-\kappa^2\left(\rho + p + \rho_{\text{eff}} + p_{\text{eff}}\right)\,.
\end{equation}
The effective fluid represents the modified part of the $f(T,B)$ Lagrangian which turns out to satisfy the conservation equation
\begin{equation}
\dot{\rho}_{\text{eff}}+3H\left(\rho_{\text{eff}}+p_{\text{eff}}\right) = 0\,.
\end{equation}

For our purpose and in order to construct the dynamical system, the $f(T,B)$ Friedmann equations can be rewritten as
\begin{eqnarray}\label{eq:friedmann_f}
&&\Omega + \Omega_{\text{eff}} = 1\,, \\
&& 3 + 2\left(\frac{H'}{H}\right) = - \frac{3f}{6H^2} + 9f_B +6f_T + 3\left(\frac{H'}{H}\right)f_B 
+ 2\left(\frac{H'}{H}\right)f_T + 2f'_T - \left(\frac{H'}{H}\right) f'_B - f''_B\,, \label{eq:friedmann_f2}
\end{eqnarray}
where
\begin{eqnarray}
\Omega_{\text{eff}} = 3f_B + 2f_T - f'_B - \frac{f}{6H^2} + \left(\frac{H'}{H}\right)f_B\,,
\end{eqnarray}
which each $i$ denotes the effective density parameter $\Omega_i = \kappa ^2 \rho_i/3H^2$. The prime $(\prime)$ denotes derivatives with respect to $N=\ln{a}$, with a chain rule given by $d/dt = H (d/dN)$.

With the latter equations we can write the continuity equations for each fluid under the consideration
\begin{eqnarray} \label{eq:conservations}
&& \rho' + 3(1 + \omega)\rho = 0, \\
&& \rho'_{\text{eff}} + 3(\rho_{\text{eff}} + p_{\text{eff}}) = 0,
\end{eqnarray}
where the effective fluid is related with the background cosmology derived from $f(T,B)$ gravity and $\omega$ are related to the cold dark matter and non-relativistic fluids as matter contributions. This set of equations impose a condition over the form of the derivative $f'(T,B)$. 

Using the Friedmann equations in Eq.(\ref{eq:friedmann_f}) and Eq.(\ref{eq:friedmann_f2}), we can directly write down the effective EoS for our $f(T,B)$ gravity as \cite{Escamilla-Rivera:2019ulu,EscamillaRivera:2010py}
\begin{eqnarray}
\omega_{\mbox{eff}} &=& \frac{p_{\mbox{eff}}}{\rho_{\mbox{eff}}}\\ 
&=& -1+\frac{\ddot{f}_B-3H\dot{f}_B-2\dot{H}f_T-2H\dot{f}_T}{3H^2\left(3f_B+2f_T\right)-3H\dot{f}_B+3\dot{H}f_B-\frac{1}{2}f}\,, \label{EoS_func}
\end{eqnarray}
which can also be written as having a redshift dependence similar to $\omega(z)$.

We can explicitly compute from Eq.(\ref{eq:friedmann_f}) a dynamical equation in terms of the Hubble factor and its derivatives as
\begin{eqnarray}
&& 6\left(\frac{H'}{H}\right)f_B + 2\left(\frac{H'}{H}\right)f_T + \left(\frac{H'^2}{H^2} + \frac{H''}{H}\right)f_B  
- \frac{f'}{6H^2} = 0\,.
\end{eqnarray}
Notice that only the last term on the r.h.s contains information about the specific form of $f(T,B)$ theory (or in its derivative).

\section{\texorpdfstring{$f(T,B)$}{f(T,B)} dynamical system structure\label{f_T_B_dynamic_struc}}

To construct the dynamical autonomous system for our $f(T,B)$ cosmological model, we follow the approach outlined in Refs.\cite{Shah:2019mxn,Mirza:2017vrk,EscamillaRivera:2010py}. As a first step we introduce
a set of conveniently specified variables which allow us to rewrite the evolution equations as an autonomous phase system. This set of equations will be subject to a generic constraint arising from our modified Friedmann equations in Eqs.(\ref{eq:friedmann_f},\ref{eq:friedmann_f2}). For this system we propose to define the parameter \cite{Odintsov:2018uaw}
\begin{equation}\label{eq:ansatz_lambda}
\lambda = \frac{\ddot{H}}{H^3} = \frac{H'^2}{H^2} + \frac{H''}{H}\,.
\end{equation}
Notice that this expression depends explicitly on $N=\ln{a}$(time-dependence). It was discussed in the latter reference that for cases when $\lambda$ = constant, some cosmological solutions can be recover, e.g. if $\lambda=0$, we can obtain a de Sitter/quasi de Sitter universe or if $\lambda =9/2$, a matter domination era can be derived. Since this ansatz shows cosmological viable scenarios as analogous to models with barotropic fluids, along the rest of this work we are going to consider $\lambda$ = constant. 
Following this prescription, we can write our Friedmann evolution equations in term of dynamical variables
\begin{eqnarray}\label{eq:aut-system}
x \equiv f_B\,, \quad
y \equiv f'_B\,, \quad
z \equiv \frac{H'}{H} = \frac{\dot{H}}{H^2}\,, \quad
w \equiv -\frac{f}{6H^2}\,.
\end{eqnarray}
From the latter definitions and the Friedmann evolution in Eq.(\ref{eq:friedmann_f}) we can derive the constriction equation from the latter evolution as
\begin{eqnarray}
    \Omega + 3x +2f_T - y +w +z x=1\,, \label{eq:constrain}
\end{eqnarray}
where $\Omega$ is a parameter that depends on the other dynamical variables. Finally, we can write the autonomous system for this theory as
\begin{eqnarray} \label{eq:variables_system}
&&z' = \lambda - 2z^2\,, \\
&&x' = y, \label{eq:variables_system2} \\
&&w' = -6zx - 2z f_T - \lambda x - 2zw\,, \label{eq:variables_system3} \\
&&y' = 3w + (9 + 3z)x + f_T(6 + 2z) + 2 f'_T - zy 
\nonumber \\ &&-3 - 2z\,. \label{eq:variables_system4}
\end{eqnarray}
To follow the constraint of the system in Eqs.(\ref{eq:friedmann_f}--\ref{eq:constrain}), we need to write $f_T$ as a dynamical variable or write it in terms of the described variables. This can be done by considering a specific form of $f(T,B)$ as we will show in section~(\ref{eq:dynamical_fTB}).

\section{Stability methodology\label{stabil_method}}
We can study our $f(T,B)$ autonomous system in Eqs.(\ref{eq:variables_system},\ref{eq:variables_system2},\ref{eq:variables_system3},\ref{eq:variables_system4}) by performing stability analyses of the critical points, which can be investigated through linear perturbations around their critical values as $\mathbf{x} = \mathbf{x}_0 + \mathbf{u}$, where $\mathbf{x}=(x,y,z,w)$ and $\mathbf{u} = (\delta x, \delta y, \delta z, \delta w)$. The equations of motion for each of our models can be written as $\mathbf{x}^\prime = \mathbf{f}(\mathbf{x})$, which upon linearisation can be given by
\begin{equation}
  \label{eq:linear}
  \mathbf{u}^\prime = \mathcal{M} \mathbf{u}\,, \quad \mathcal{M}_{ij}
  = \left. \frac{\partial f_i}{\partial x_j}
  \right|_{\mathbf{x}_\ast}\,, 
\end{equation}
where $\mathcal{M}$ is known as the linearisation matrix \cite{Bahamonde:2017ize}. The eigenvalues indicated by $\omega$ of $\mathcal{M}$ determine the stability (type) of the critical points, whereas the eigenvectors $\mathbf{\eta}$ of $\mathcal{M}$ indicates the principal directions of the perturbations performed at linear level. As it is standard in the stability analysis, if $\mathrm{Re}(\omega) < 0$
($\mathrm{Re}(\omega) > 0$) the critical point is called stable (unstable). More specific types of point will be indicated for each $f(T,B)$ scenario.

For this case, we should consider perturbations of the four dynamical variables $(x,y,z,w)$, keeping in mind that they are not all completely independent because they are bound together by the Friedmann constraint in Eq.(\ref{eq:friedmann_f}). This dependence would then carry over to perturbative level.

From Eq.(\ref{eq:variables_system}) we notice that there is not an explicit dependency of $f(T,B)$, therefore for the critical points following the above prescription $\mathbf{x_*}$ / $\mathbf{\dot{x}} = \mathbf{f}(\mathbf{x_*})=0$ we require that
 \begin{equation}
    z = \pm \sqrt{\frac{\lambda}{2}}\,, ~ \quad ~
    y=0\,.
\end{equation}
For each $f(T,B)$ cosmological case we will present the stability results, where we only consider the eigenvalues of the stability matrix $\mathcal{M}$ for each of the critical points and for the perturbations that are compatible with Eq.(\ref{eq:friedmann_f}).

\section{Dynamical analyses for \texorpdfstring{$f(T,B)$}{f(T,B)} models \label{dynami_ana}}
\label{eq:dynamical_fTB}
In this work we consider three $f(T,B)$ scenarios. They were selected in order to obtain cosmological viable cases, in particular the late-time observed cosmic acceleration. The following models were studied in detail in Ref.\cite{Escamilla-Rivera:2019ulu}, where cosmological constraints of each of them were found. In the following, we will focus on the corresponding parameter values which adapt to our autonomous system. 

\subsection{Stability analysis for General Taylor Expansion model}
\begin{widetext}
\begin{figure*}
\centering
      \includegraphics[width= 0.31\textwidth]{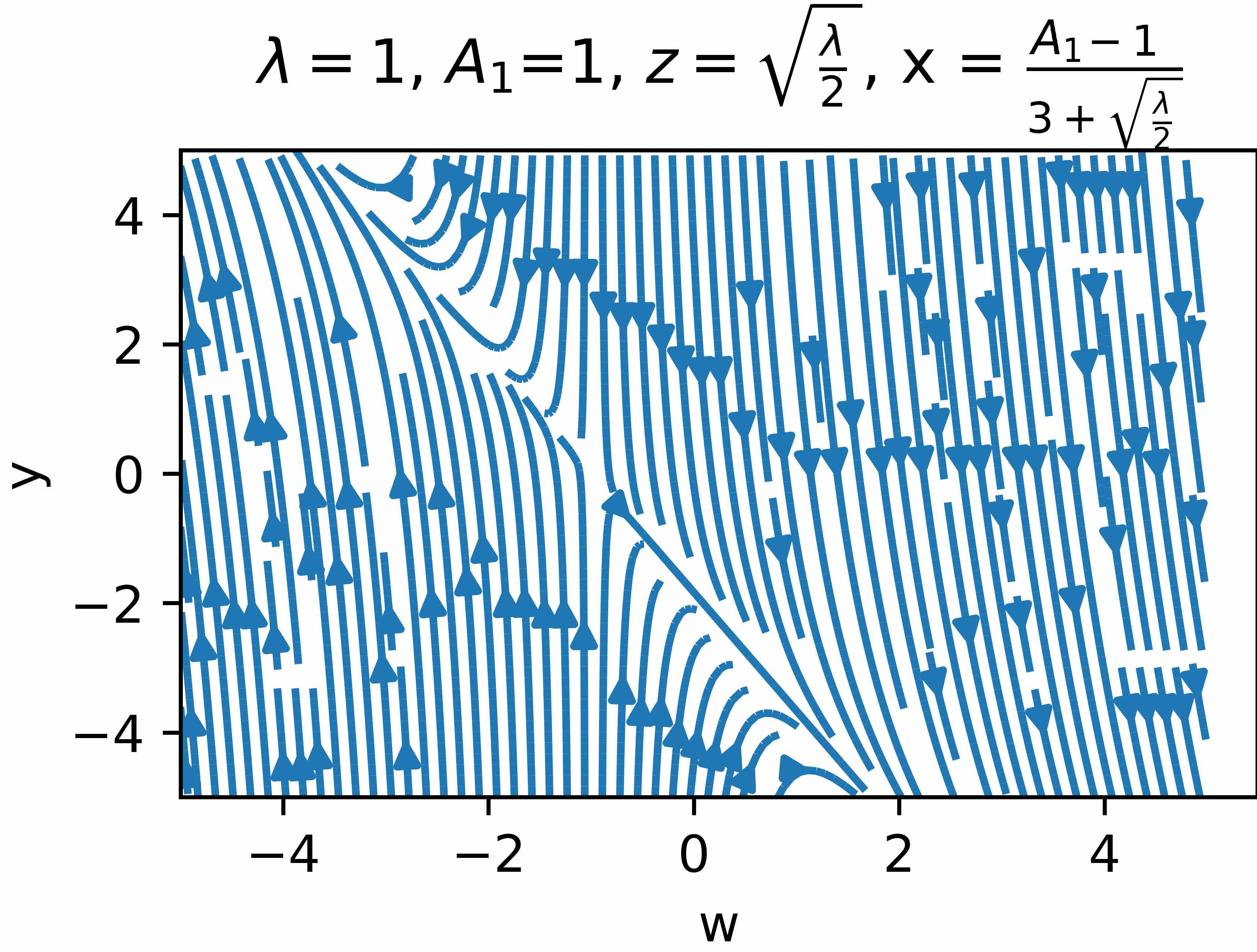}
      \includegraphics[width= 0.31\textwidth]{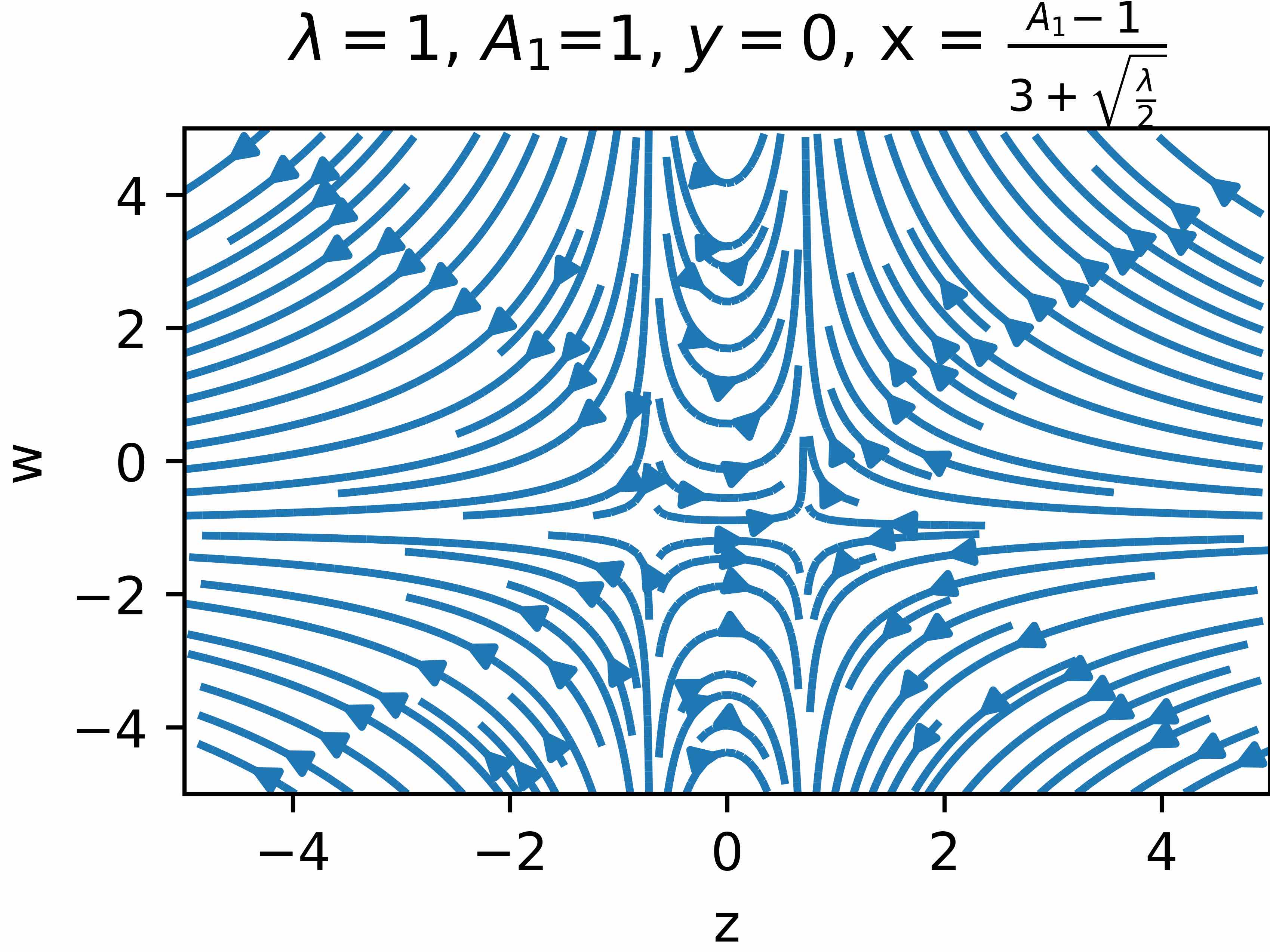}
       \includegraphics[width= 0.31\textwidth]{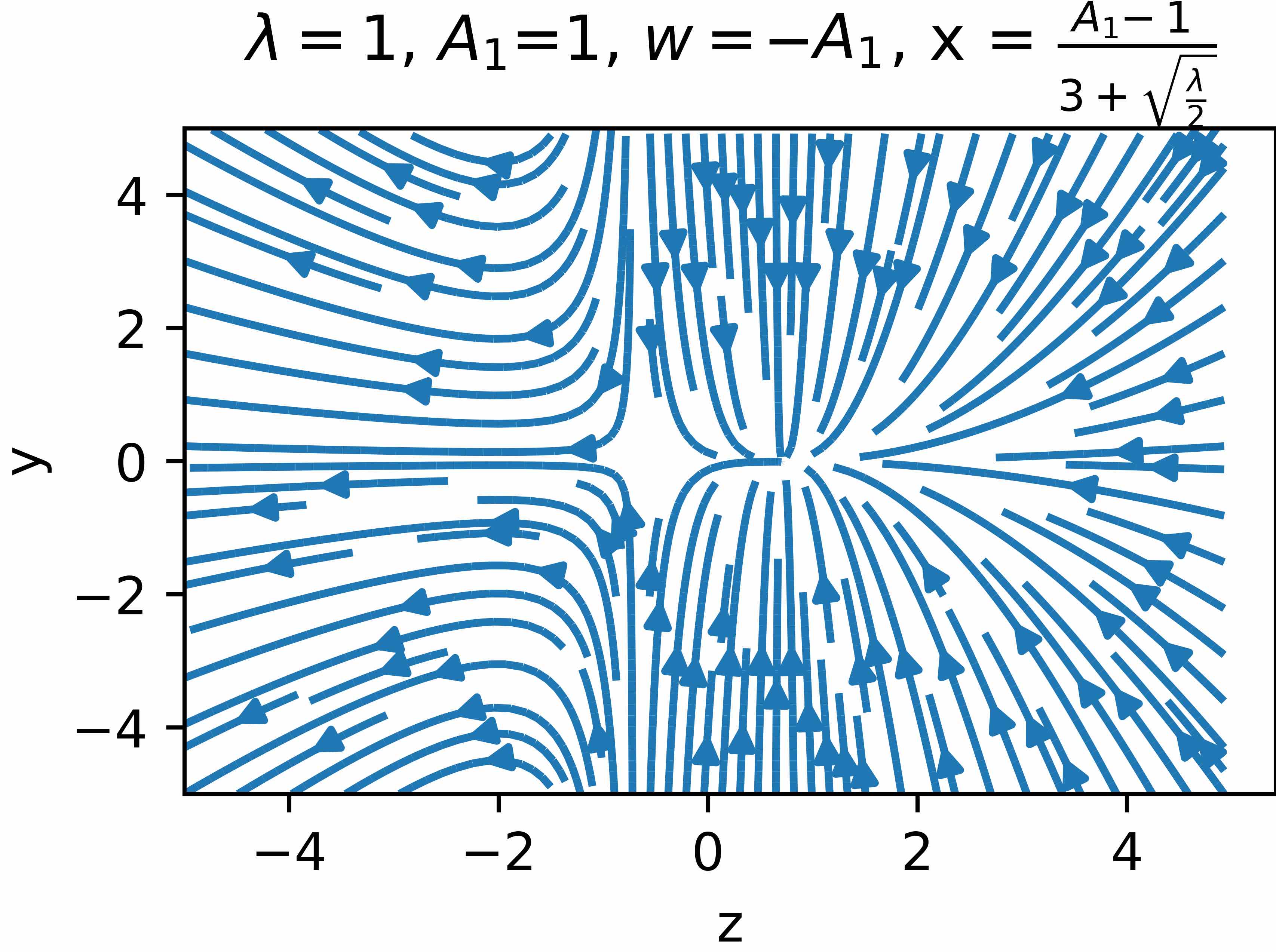}
      \includegraphics[width= 0.31\textwidth]{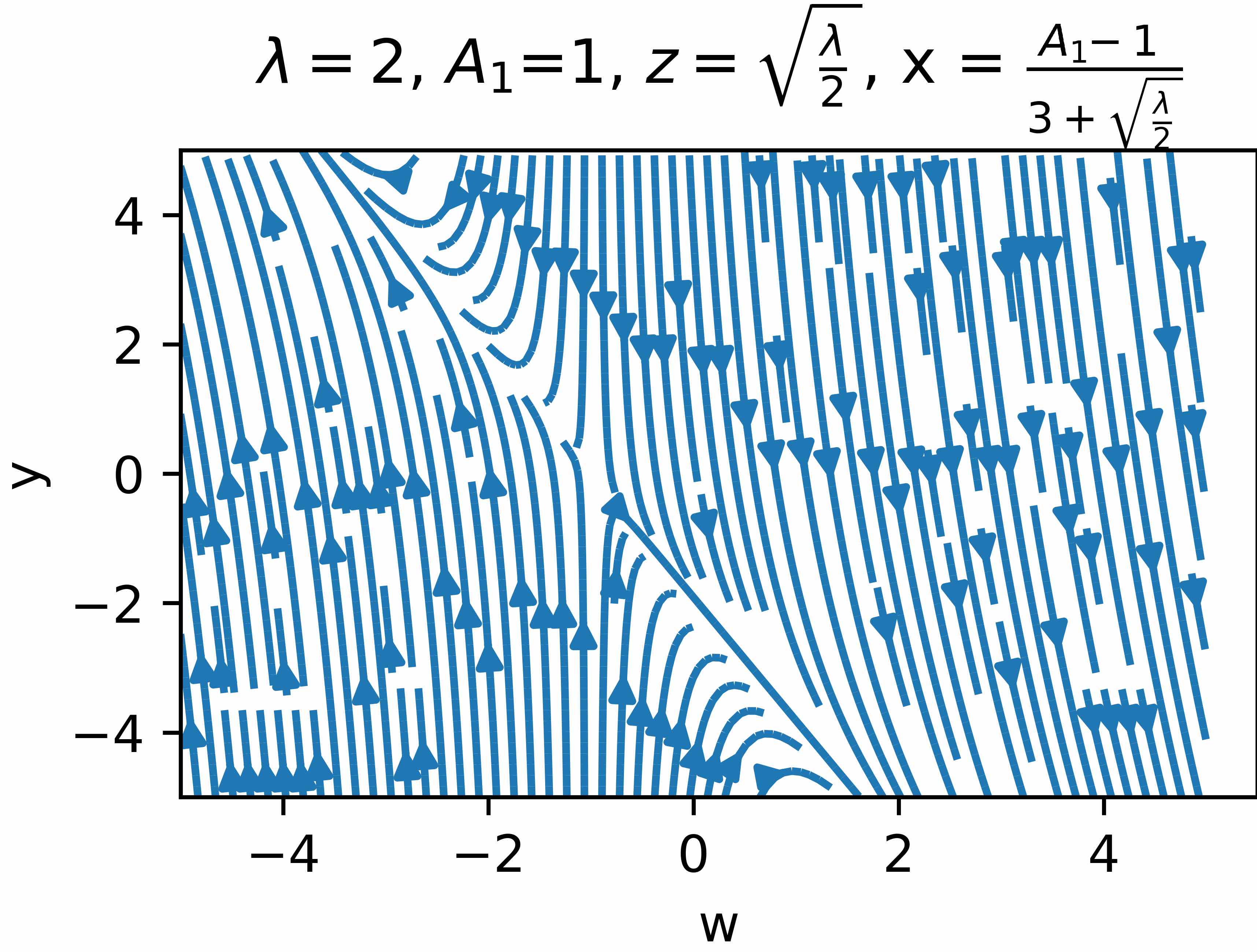}  
         \includegraphics[width= 0.31\textwidth]{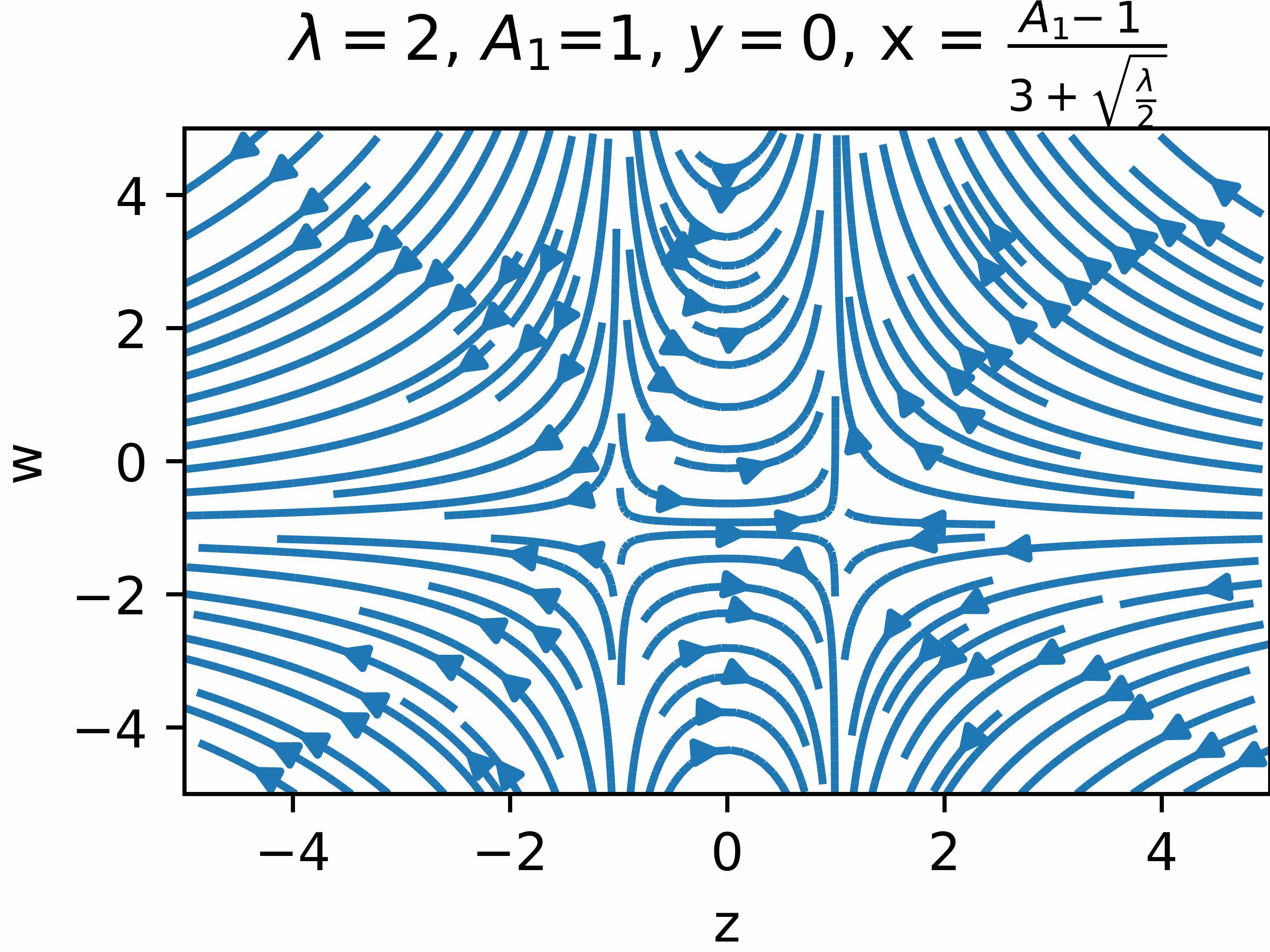}
\includegraphics[width= 0.32\textwidth]{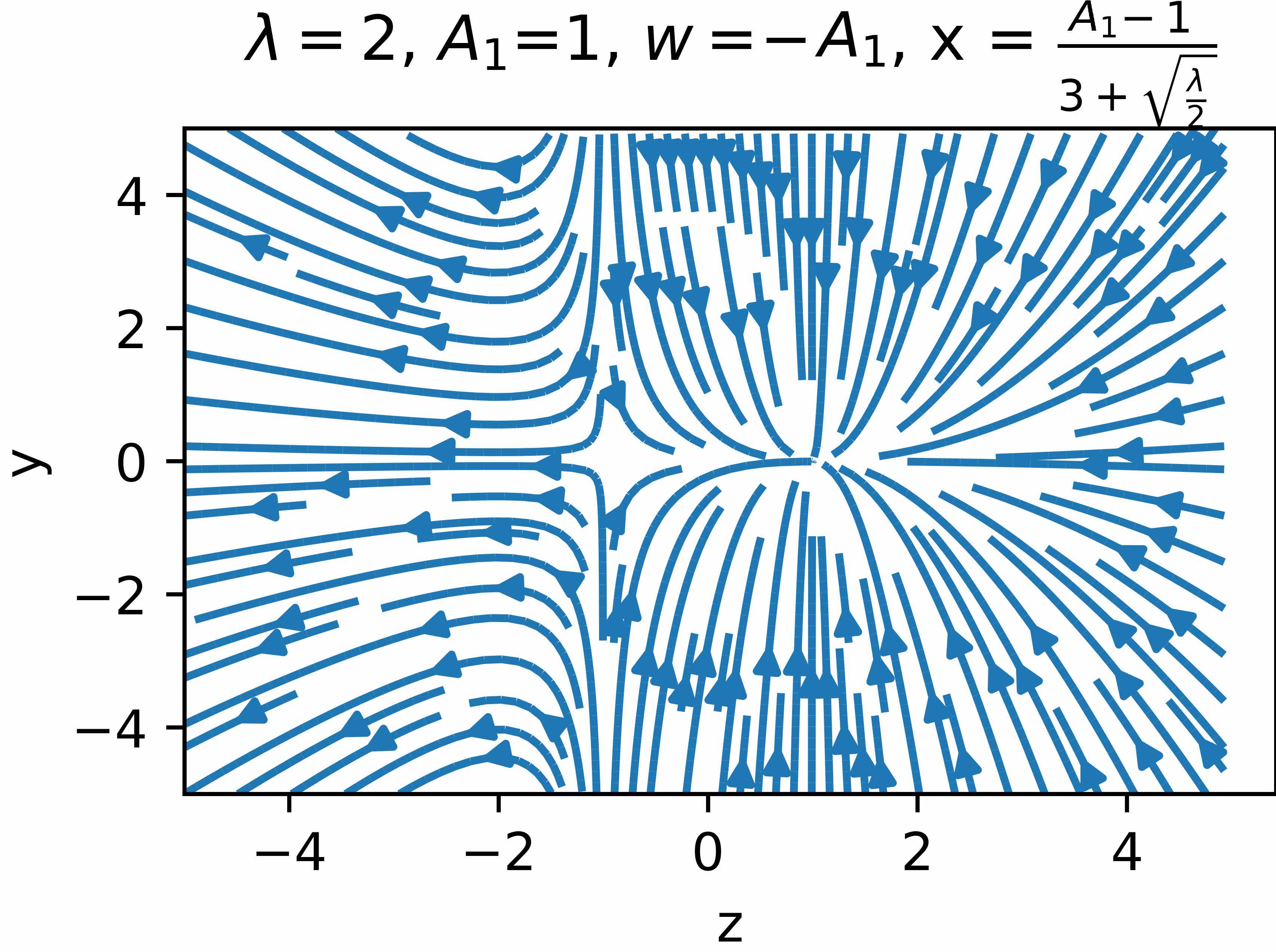}
    \includegraphics[width= 0.31\textwidth]{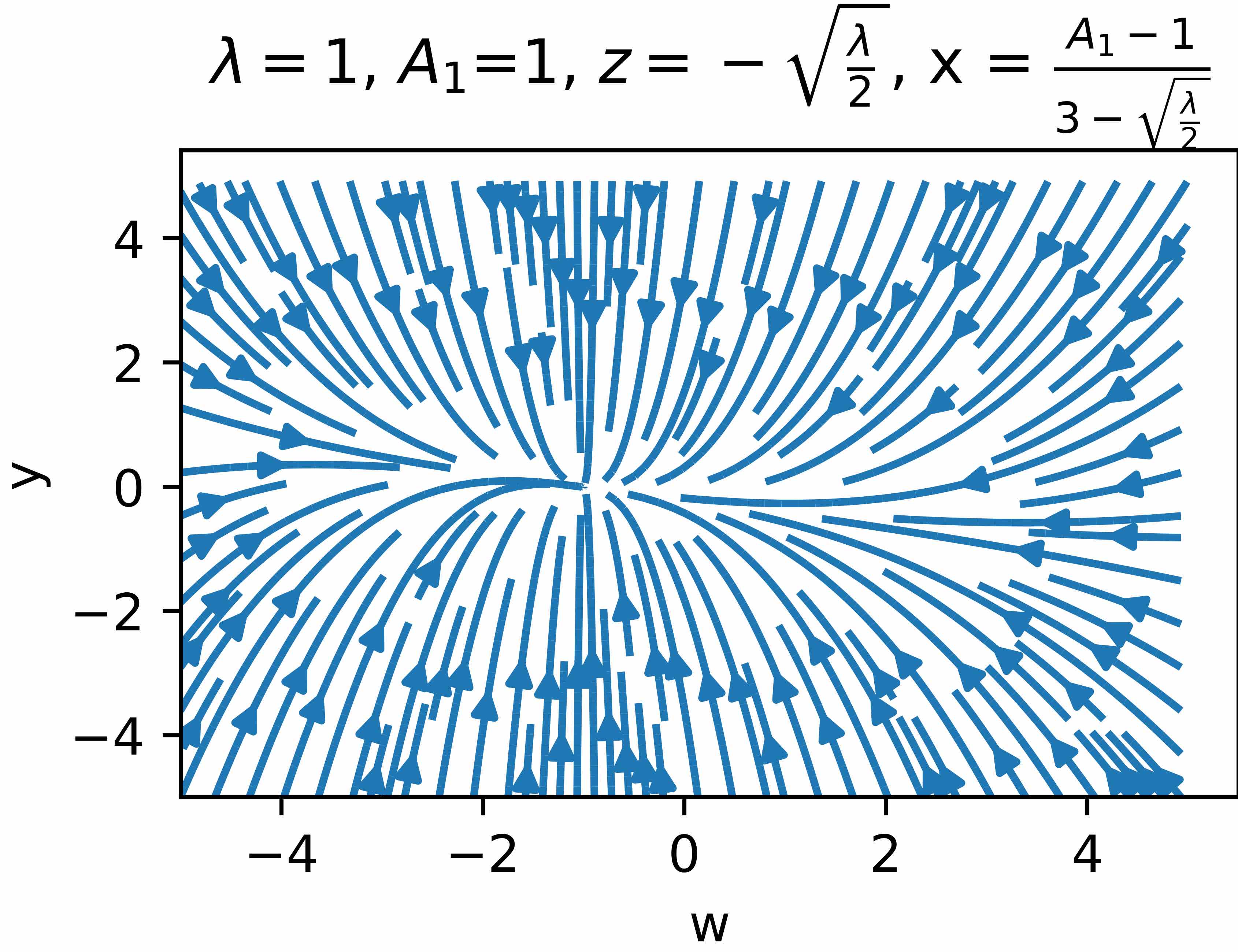}
    \includegraphics[width= 0.32\textwidth]{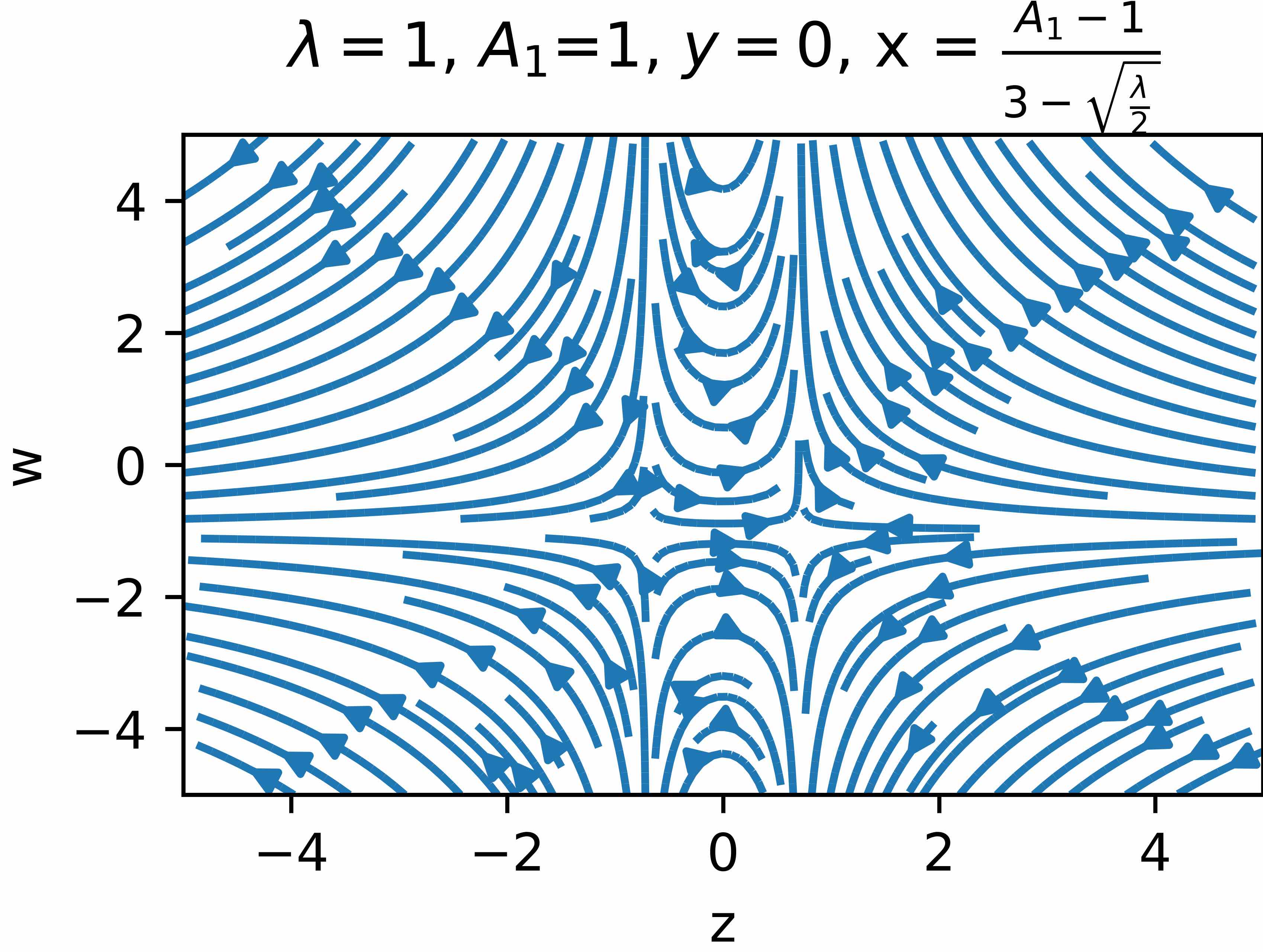}
    \includegraphics[width= 0.31\textwidth]{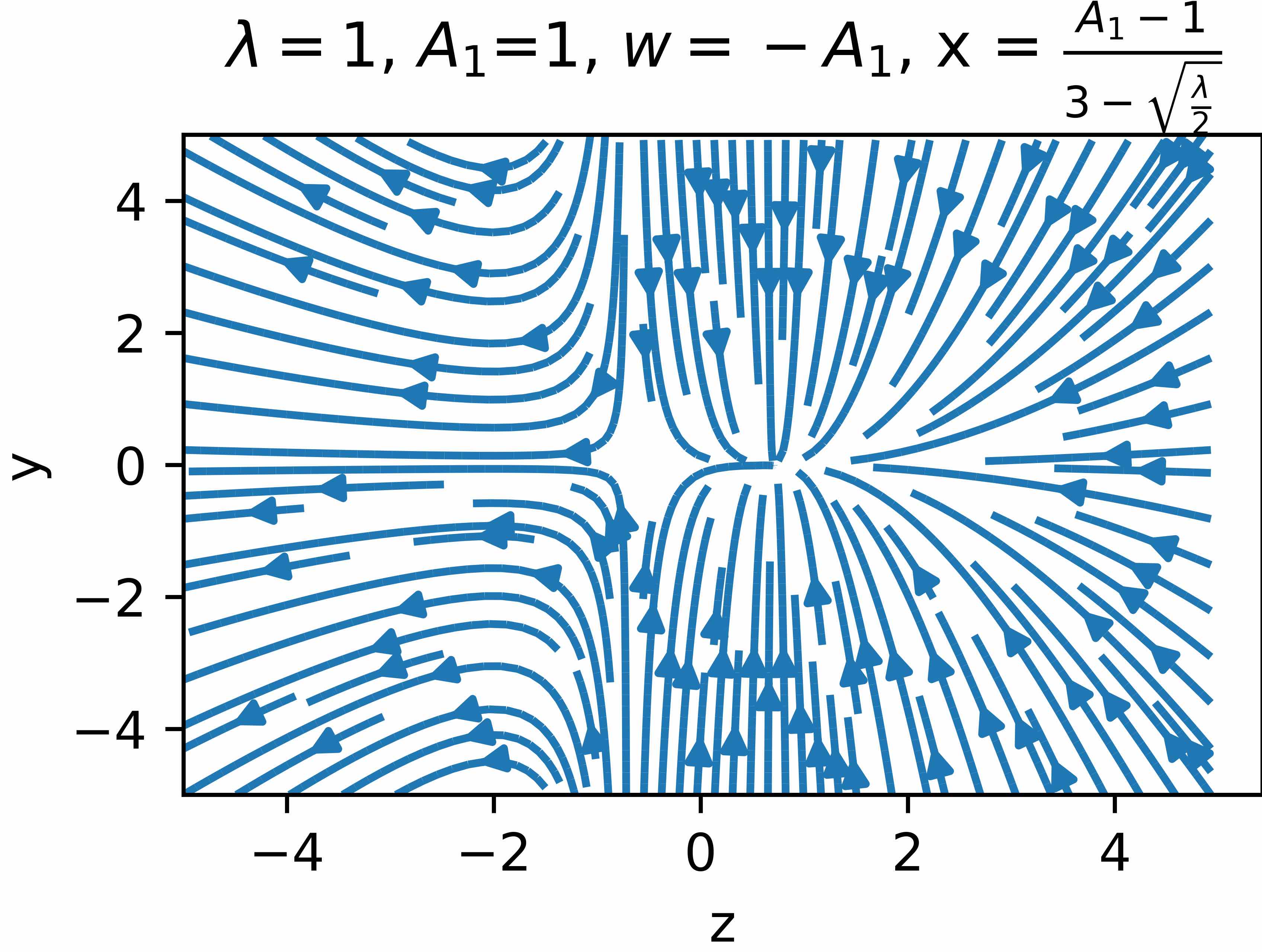}
    \includegraphics[width= 0.31\textwidth]{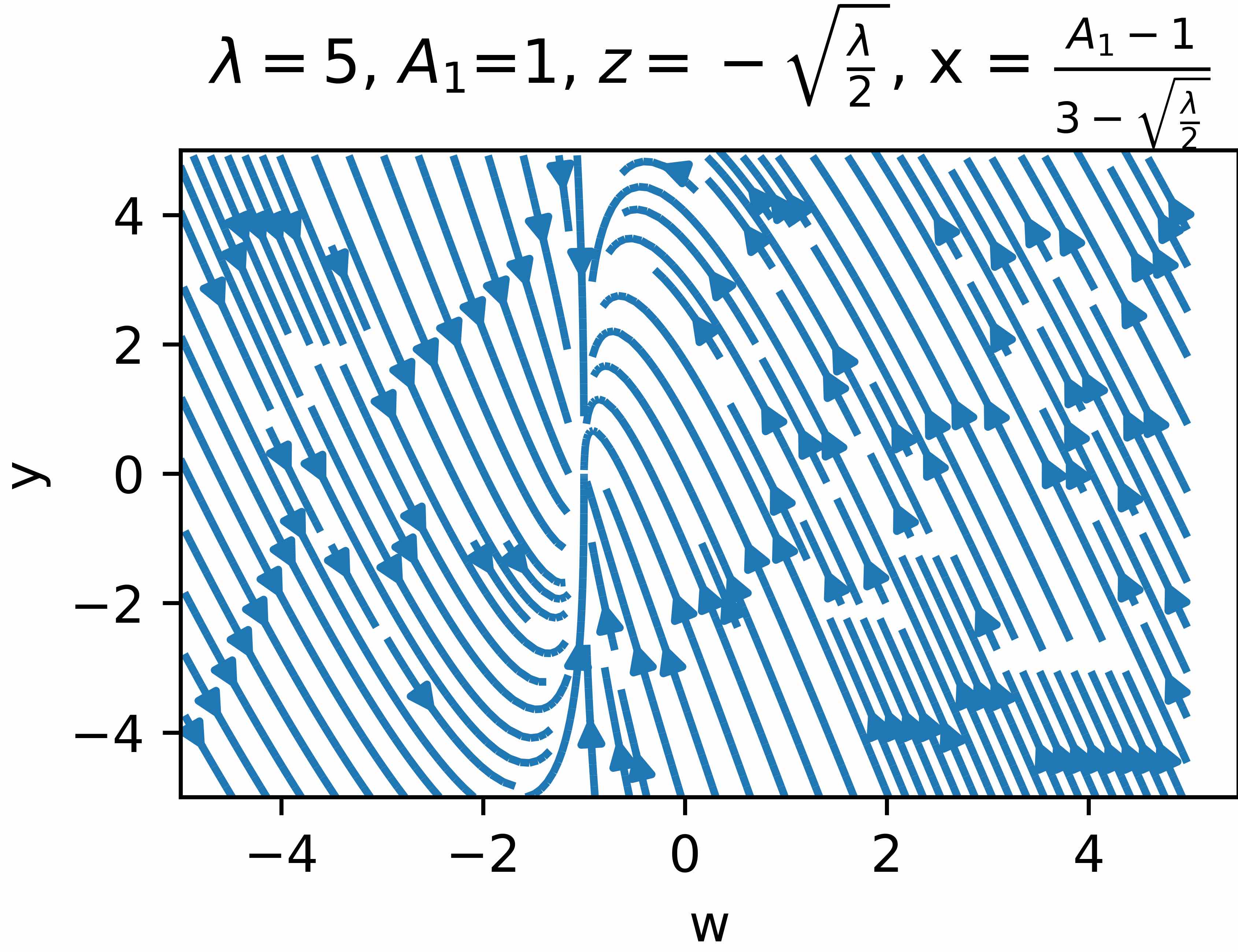}
    \includegraphics[width= 0.31\textwidth]{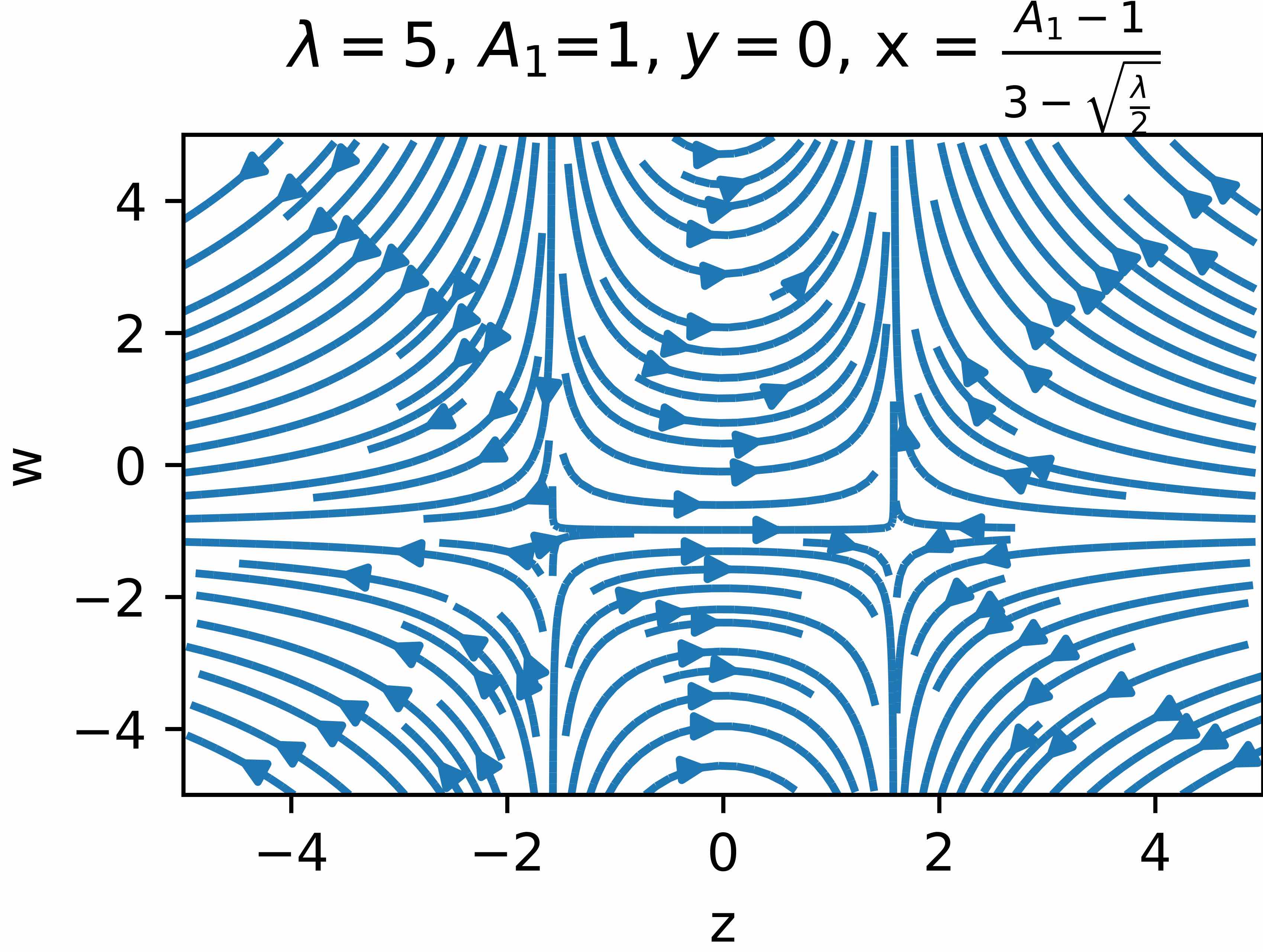}
    \includegraphics[width= 0.31\textwidth]{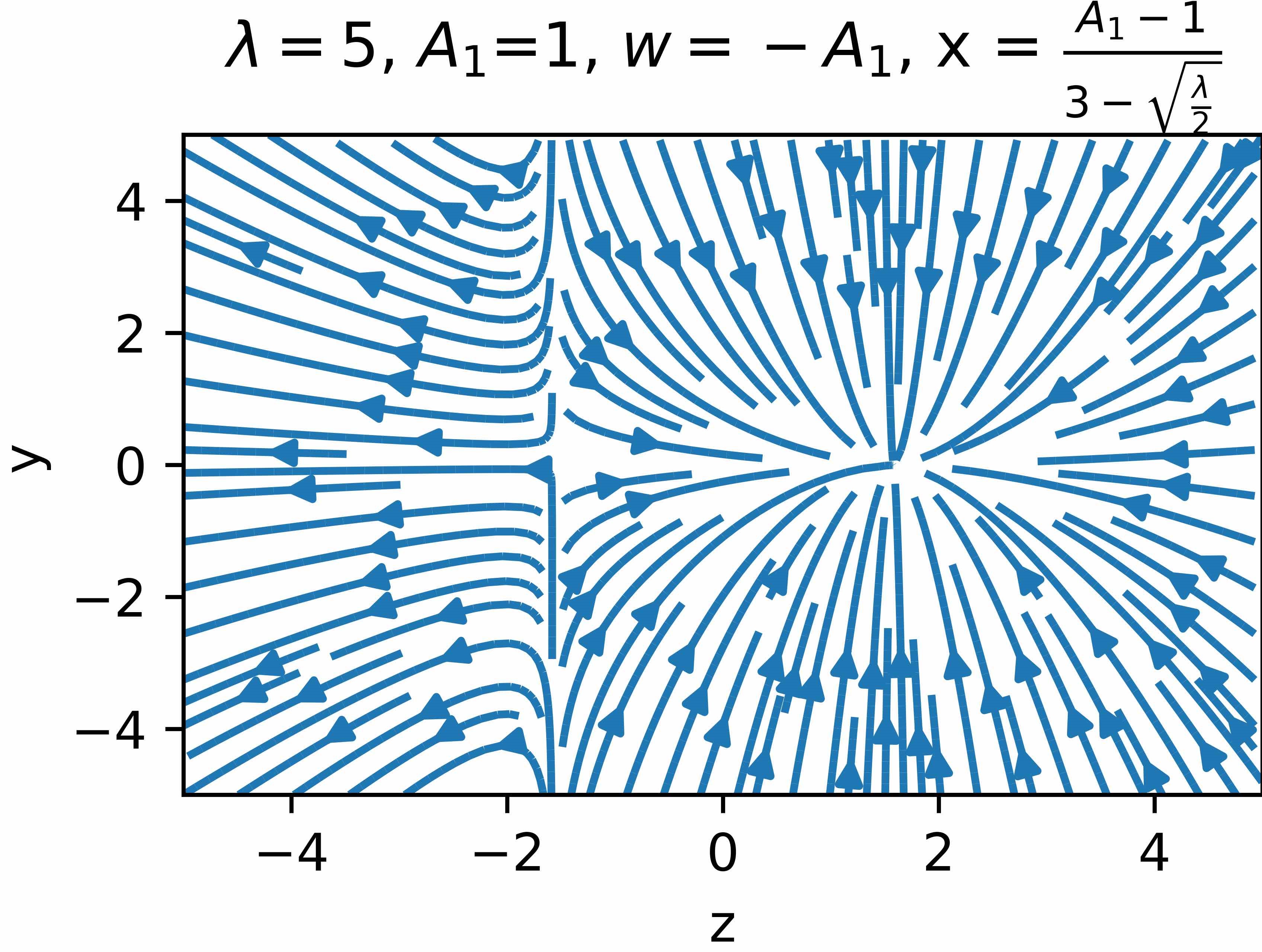}
          \caption{Different views of the phase space of the dynamical system in Eq.(\ref{eq:taylor_system}) for $\lambda \neq 0$. The system was reduced to a 2-d surface representing the different perspectives of its constraint. The arrows represent the direction of the velocity field and the trajectories reveal their stability properties as described for this model.}
    \label{fig:taylor_case}
\end{figure*}
\end{widetext}

The form for this model was presented in Ref.\cite{Farrugia:2018gyz} as a general Taylor expansion of the $f(T,B)$ Lagrangian, given by
\begin{eqnarray}\label{taylor_model}
&& f(T,B) = f(T_0, B_0) + f_T(T_0,B_0) (T-T_0) 
 + f_B(T_0,B_0) (B-B_0) + \frac{1}{2!}f_{TT}(T_0,B_0) (T-T_0)^2  \nonumber\\
 &&+ \frac{1}{2!}f_{BB}(T_0,B_0) (B-B_0)^2  
  + f_{TB}(T_0,B_0) (T-T_0)(B-B_0) + \mathcal{O}(T^3,B^3)\,,
\end{eqnarray}
which gives the general Taylor expansion of the $f(T,B)$ Lagrangian about its Minkowski values for the torsion scalar $T$ and boundary term $B$. We notice from here that we need to take into account beyond linear approximations since $B$ is a boundary term at linear order. Following the form for the FLRW tetrad in Eq.(\ref{FLRW_tetrad}), where locally spacetime appears to be Minkowski with torsion scalar and boundary term values, we can consider $ T_0 = B_0 = 0\,$. Taking constants called $A_i$, the Lagrangian can be rewritten as
\begin{equation}
f(T,B)\simeq A_0+A_1 T + A_2 T^2 + A_3 B^2 + A_4 TB\,,
\end{equation}
where the linear boundary term vanishes. We notice from this specific form that the first term can be seen as $A_0 \approx \Lambda$, therefore we are dealing with a cosmic acceleration as a consequence of the $f(T,B)$. Thus, the form of this model can be written in terms of the dynamical variables as
\begin{equation} \label{eq:taylor_system}
    f_T = -(3+z)x - 2w - A_1\,,
\end{equation}
at linear order in torsion and with $A_0=0$, i.e we are switching-off the cosmological constant. This can be done since an explicitly time-dependent factor appears and then a different approach has to be taken. The critical points for this model are 
\begin{eqnarray}\label{taylor_crit}
    w= -A_1\,, ~ \quad ~
    x = \frac{A_1 -1}{3 \pm \sqrt{\frac{\lambda}{2}}}\,,
\end{eqnarray} 
which imply that the constriction evolution equation in Eq.\eqref{eq:constrain} is now $\Omega=0$. According to these points we can compute the following eigenvalues for the system
\begin{eqnarray}\label{taylor_eigenvalues}
\omega_{1} &=&-3 \mp \sqrt{\frac{\lambda}{2}}\,, \quad \omega_{2} =-3 \mp 2\sqrt{\frac{\lambda}{2}}\,, 
\nonumber \\
\omega_{3} &=&\mp 4\sqrt{\frac{\lambda}{2}}\,, \quad  \omega_{4} =\pm 2\sqrt{\frac{\lambda}{2}}\,.
\end{eqnarray}
\\
Considering values as $\lambda \neq 0$, we get that $Re(\omega_3) = - Re(\omega_4)\neq 0$, implying that for this system all the critical points are saddle-like for any value of $\lambda$. In Fig.(\ref{fig:taylor_case}), we show different views of the phase space of the dynamical system in Eq.(\ref{eq:taylor_system}) on 2-d surfaces. The solutions for the case are in agreement with the cosmological constraints found in \cite{Escamilla-Rivera:2019ulu}. According to these results, our critical points behave as $A_{i}< A_{i+1}$ (which states for these values $A_0=0$ and $A_1=1$), show a quintessence behaviour and when $B$ dominates and $z\approx 1$. After that, a $\Lambda$CDM model behaviour is observed. 

\subsection{Stability analysis for Power Law model \label{power_law_sec}}
If we consider a Lagrangian of separated power law style models for the torsion and boundary scalars, we can write a model like \cite{Bahamonde:2016grb} 
\begin{equation}
    f(T,B) = b_0 B^k + t_0 T^m\,. \label{eq:powerlaw}
\end{equation}
This is an interesting model since it was already been shown in Ref.\cite{Said:2017nti} that for $m<0$ the Friedmann equations will be effected mostly in the accelerating late-time universe while for $m>0$ this impact will take place for the early universe, assuming no input from the boundary contribution. By incorporating the boundary term, this analysis will reveal an effect of $B$ on the combined evolution within $f(T,B)$ cosmology. The form for this model can be written in terms of the dynamical variables as:
\begin{equation}\label{eq:powerlaw_dynamical}
    f_T = -mw - \frac{m}{k}(3+z)x\,.
\end{equation}
The critical points for this scenario are
\begin{eqnarray}
    w = \frac{k-m}{m(1-k)}\,, ~ \quad ~
    x = -\frac{k}{m} \Big(\frac{m-1}{k-1}\Big) \frac{1}{3 \pm \sqrt{\frac{\lambda}{2}}}\,.
\end{eqnarray}
For this case, the constriction \eqref{eq:constrain} gives again $\Omega=0$.

According to Eq.(\ref{eq:powerlaw_dynamical}), we can analyse independently the positive and negative roots with $z = \pm \sqrt{\frac{\lambda}{2}}$ as follows.

\subsubsection{Analysis for the positive branch}

\paragraph{Critical points.}

For this case, the eigenvalues derived from the stability matrix are
\begin{eqnarray}
\tiny
\omega_1 &=& -\sqrt{2} \sqrt{\lambda }-3\,, \\
\omega_2 &=& -2 \sqrt{2} \sqrt{\lambda }\,, \\
\omega_3 &=& -\frac{1}{4 k}\left(\sqrt{\alpha -\beta +\gamma}
    +k\left(\sqrt{2} \sqrt{\lambda } (1-2 m)-6\right)+2 \left(\sqrt{2} \sqrt{\lambda}+6\right) m \right)\,,\\
\omega_4 &=& \frac{1}{4 k}\left(\sqrt{\alpha -\beta +\gamma}
+k
   \left(\sqrt{2} \sqrt{\lambda } (2 m-1)+6\right)-2 \left(\sqrt{2} \sqrt{\lambda
   }+6\right) m \right)\,,
\end{eqnarray}
where 
\begin{eqnarray}
\alpha &=& 2 k^2 \left(\lambda  (2 m+1)^2+6 \sqrt{2} \sqrt{\lambda } (6m-1)+18\right)\,, \\
\beta&=&8 k m \left(\lambda +6 \sqrt{2} \sqrt{\lambda } (m+1)+2 \lambda m+18\right)\,,\\
\gamma&=&8 \left(\lambda +6 \sqrt{2} \sqrt{\lambda }+18\right) m^2\,,
\end{eqnarray}
which under the conditions  $Re(\omega_1), Re(\omega_2) < 0$ for any value of $\lambda$ we get saddle/attractors points. 

\begin{itemize}
\item Case $Re(\omega_3) = 0$.
For the condition $Re(\omega_3) = 0$ the critical regions are\footnote{From this point, along the text we refer to the symbol $\lor$ as \textit{or}, $\land$ as \textit{and}.}
\begin{enumerate}
\item \begin{align}
k=1\land 0<m<\frac{1}{2}\land 0<\lambda <72 m^2-72 m+18\,,
\end{align}
\item \begin{align}
0<m<\frac{1}{2}\land 2 m<k\leq 1\land \lambda =\frac{18 (k-2 m)^2}{(-2 k m+k+2 m)^2}\,,
\end{align}
\end{enumerate}

\item For the condition $Re(\omega_4) = 0$ the critical regions are  
\begin{enumerate}
\item \begin{align}
k=1\land \left[\left(0<m<\frac{1}{2}\land \lambda >72 m^2-72 m+18\right)\lor \left(m\geq \frac{1}{2}\land \lambda >0\right)\right]\,,
\end{align}
\item \begin{align}
0<m<\frac{1}{2}\land 2 m<k\leq 1\land \lambda =\frac{18 (k-2 m)^2}{(-2 k m+k+2 m)^2}\,,
\end{align}
\end{enumerate}

\item Attractor regions. These cases can happen under the following conditions:
\begin{enumerate}
\item \begin{align}
0 < m \leq \frac{1}{2} \land 0 < k < 1 \land \lambda > 18 \,,
\end{align}
\item \begin{align}
m > \frac{1}{2} \land  0 < k < 1 \land \lambda > 0 \,,
\end{align}
\end{enumerate}
\end{itemize}

\paragraph{Properties:} 
\begin{itemize}
\item If $m> k$ then, $w<-1/3$ ($b_0$ and $c_0$ fixed as positive).
\item If $m< k$ then, we get $\Lambda$CDM. ($b_0$ and $c_0$ fixed as positive).
\item If $b_0 < t_0$ and vice versa, we get a crossover over the phantom divided-line ($w=-1$).
\item We recover $\Lambda$CDM and late cosmic acceleration.
\end{itemize}


\subsubsection{Analysis for the negative branch} 

\paragraph{Critical points.}
For this case, the eigenvalues derived from the stability matrix are given by
\begin{align}
\tiny
\omega_1 &= \sqrt{2} \sqrt{\lambda }-3\,, \\
\omega_2 &= 2 \sqrt{2} \sqrt{\lambda }\,, \\
\omega_3 &= \frac{1}{4 k}\left(-\sqrt{\frac{\gamma k^2}{4 m^2}+\eta}\right) 
+\zeta\,, \\
   \omega_4 &= \frac{1}{4 k}\left(\sqrt{\frac{\gamma k^2}{4 m^2}+\eta}\right)
   +\zeta\,.
\end{align}
with
\begin{eqnarray}
\zeta&=& \frac{1}{4 k}\left(k \left(\sqrt{2} \sqrt{\lambda } (1-2 m)+6\right)+2 \left(\sqrt{2} \sqrt{\lambda }-6\right) m\right)\,,\\ 
\eta &=& 8 m^2
   \left(\lambda  (k-1)^2+6 \sqrt{2} \sqrt{\lambda } (k-1)+18\right)- 8 m k
   \left(3 \sqrt{2} \sqrt{\lambda } (3 k-2)-k \lambda +\lambda +18\right)\,.
\end{eqnarray}
Notice that according to the values of $\omega_1$ and $\omega_2$, the critical point associated to the negative root case corresponds to a scenario where the universe has a contraction (accelerated) phase $\lambda < 2$($\lambda >2$), respectively, which represents a saddle point. On the other hand, according to the value of $\omega_1$, we notice that if $0 <\lambda < \frac{9}{2}$, the critical point is hyperbolic and saddle-type. To simplify the analysed regions, we consider cases where $\omega_1$, or $\omega_2$, only have a non-vanishing real part. The next case to explore will be with a vanished real part (which correspond to a non-hyperbolic case).

\begin{itemize}
\item Conditions with $Re(\omega_3) = 0$. These regions are
\begin{enumerate}[label=\Alph*:]
\item \begin{align}
 0<m < 1\land \left((0<k<1\land \lambda =18)\lor \left(k=1\land \lambda >72 m^2-72 m+18\right)\lor (k>1\land \lambda =18)\right),
\end{align}
\item \begin{align}
k=1\land \left[\left(m\geq 1\land\lambda \geq 72 m^2-72 m+18\right)\lor \left(\frac{3}{4}<m<1\land
   \lambda =72 m^2-72 m+18\right)\right]\,,
\end{align}
\item \begin{align}
0<k<1\land \left(m\geq1\land \lambda =\frac{18 (k-2 m)^2}{(-2 k m+k+2 m)^2}\right),
\end{align}
\item \begin{align}
1<k<2\land \left(\frac{3 k}{2 k+2}<m\leq 1 \land \lambda =\frac{18 (k-2 m)^2}{(-2 k m+k+2 m)^2}\right) \,,
\end{align}
\item \begin{align}
k=2\land m=1\land \frac{9}{2}<\lambda \leq 18\,,
\end{align}
\item \begin{align}
k>2\land 1\leq m<\frac{3 k}{2 k+2}\land \lambda =\frac{18 (k-2 m)^2}{(-2 k m+k+2 m)^2}\,,
\end{align}
\item \begin{align}
k<0\land \left[(m>1\land \lambda =18)\lor \left(1\leq m\leq \frac{3}{2}\land \lambda =\frac{18 (k-2 m)^2}{(-2 k m+k+2 m)^2}\right)\right]\,,
\end{align}
\item \begin{align}
m>\frac{3}{2}\land -\frac{2 m}{2 m-3}<k<0\land \lambda =\frac{18 (k-2 m)^2}{(-2 k m+k+2 m)^2}\,,
\end{align}
\end{enumerate}

\item Condition  $Re(\omega_4) = 0$. These regions are:
\begin{enumerate}[label=\Alph*:]
\item \begin{align}
k=1\land \frac{3}{4}<m\leq 1\land \frac{9}{2}<\lambda \leq 72 m^2-72 m+18\,,
\end{align}
\item \begin{align}
\textstyle
&\left[m\geq 1\land 0<k<1\land \left(\lambda =18\lor \lambda =\frac{18 (k-2 m)^2}{(-2 k m+k+2 m)^2}\right)\right]
\nonumber\\&
\lor \left(m>1\land k=1\land \frac{9}{2}<\lambda <72 m^2-72 m+18\right) \nonumber\\&
\lor (k>1\land\lambda =18)\,,
\end{align}
\item \begin{align}
1<k<2\land \left[\left(\frac{3 k}{2 k+2}<m<1\land \lambda =\frac{18 (k-2 m)^2}{(-2 k m+k+2 m)^2}\right)\lor (m=1\land \lambda =18)\right]\,,
\end{align}
\item \begin{align}
k=2\land m=1\land \frac{9}{2}<\lambda \leq 18\,,
\end{align}
\item \begin{align}
k>2\land \left(1\leq m<\frac{3 k}{2 k+2}\land \lambda =\frac{18 (k-2 m)^2}{(-2 k m+k+2 m)^2}\right),
\end{align}
\item \begin{align}
k<0\land \left[(0<m<1\land \lambda =18)\lor \left(1\leq m\leq \frac{3}{2}\land \lambda =\frac{18 (k-2 m)^2}{(-2 k m+k+2 m)^2}\right)\right]\,,
\end{align}
\item \begin{align}
m>\frac{3}{2}\land -\frac{2 m}{2 m-3}<k<0\land \lambda =\frac{18 (k-2 m)^2}{(-2 k m+k+2 m)^2}\,.
\end{align}
\end{enumerate}

\item Saddle regions. These regions are determine by the condition $Re(\omega_3),Re(\omega_4) >0$, therefore
\begin{enumerate}[label=\Alph*:]
\item \begin{align}
0<m<\frac{1}{2} \land\left[ \left(0 < k < 1 \land \lambda >18 \right)\lor \left(k>1 \land \frac{9}{2}< \lambda < 18\right)\right]\,,
\end{align}
\item \begin{align}
m=\frac{1}{2} \land \left[  \left( 0<k < 1 \land \lambda >18 \right)\lor \left( k>1 \land \frac{9}{2}<\lambda < 18\right) \right]\,,
\end{align}
\item \begin{align}
\frac{1}{2} < m < 1 \land 0 < k < 1 \land \lambda > 18\,,
\end{align}
\item \begin{align}
m \geq 1 \land \left[\left( 0 < k < \frac{2m}{1+2m} \land \lambda > 18 \right) \lor \left( k > \frac{2m}{1+2m} \land \lambda > 18 \right) \right]\,,
\end{align}
\item \begin{align}
1<m<\frac{3}{2}\land k>-\frac{2 m}{2 m-3}\land \frac{9}{2}<\lambda <\frac{18 (k-2 m)^2}{(-2 k m+k+2 m)^2}\,,
\end{align}
\item \begin{align}
k <0 \land \left[\left( 0 < m < 1 \land \frac{9}{2} < \lambda < 18 \right) \land \left( m > 1 \land \frac{9}{2} < \lambda < \frac{18(k-2m)^2}{(k+2m -2km)^2}\right) \right]\,.
\end{align}
\end{enumerate}
\end{itemize}

\begin{widetext}
\begin{figure*}
\centering
   \includegraphics[width= 0.31\textwidth]{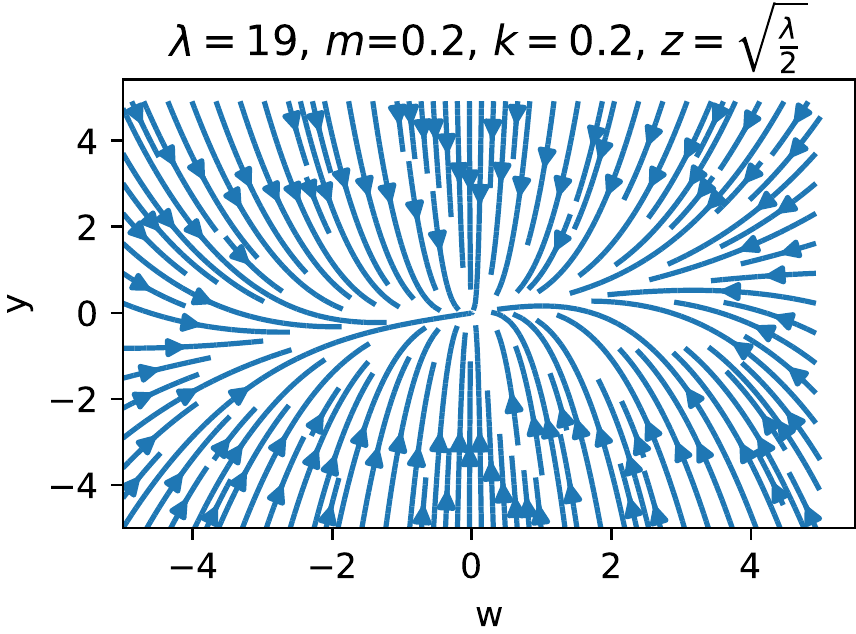}
      \includegraphics[width= 0.31\textwidth]{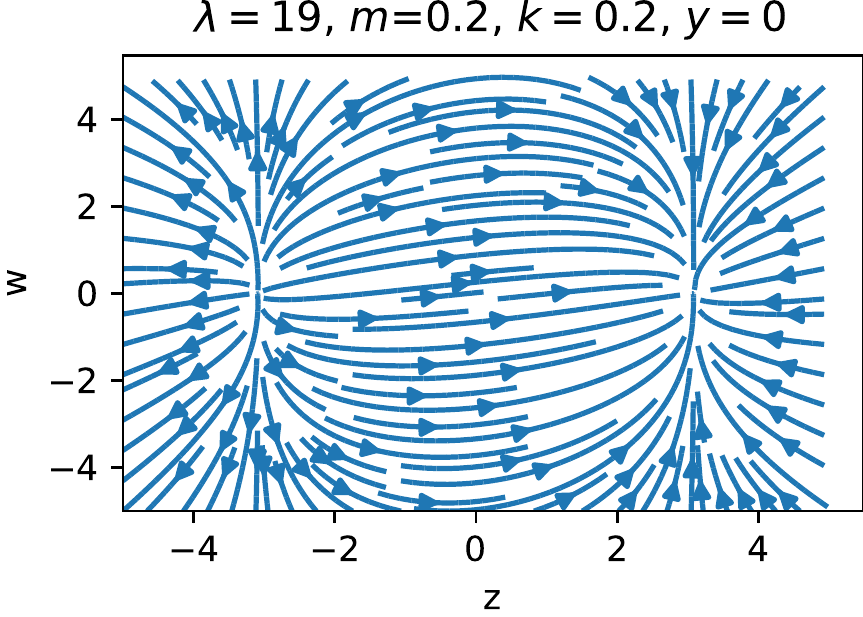}
      \includegraphics[width= 0.31\textwidth]{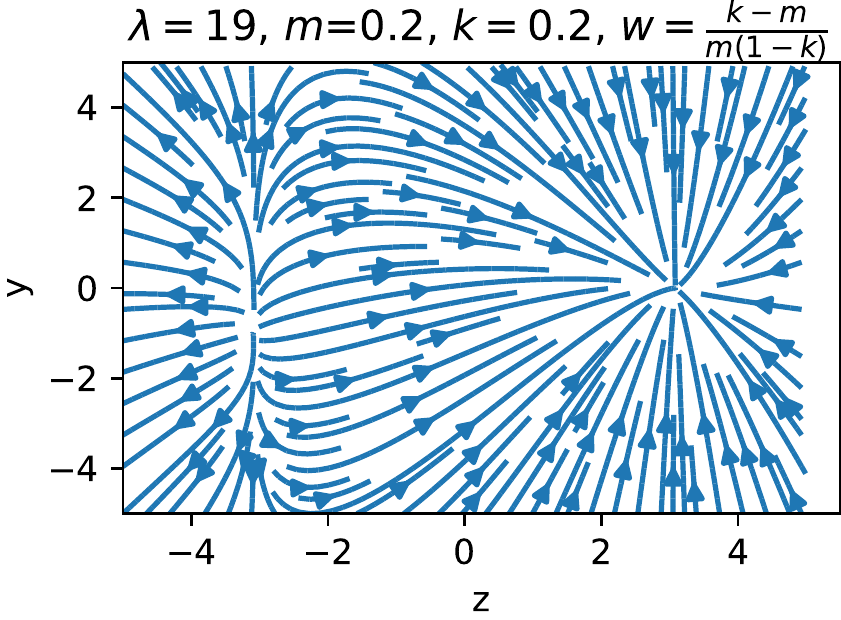}
    \includegraphics[width= 0.31\textwidth]{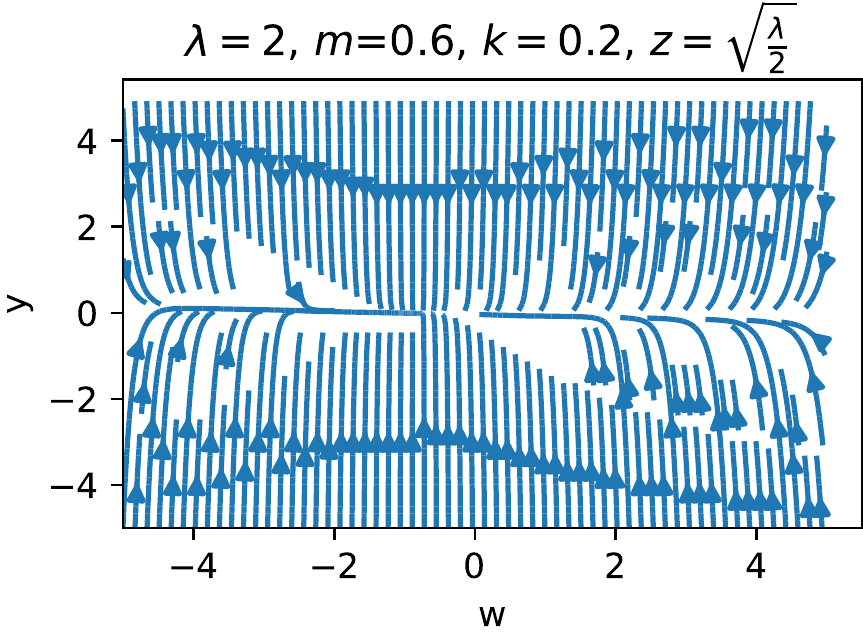}
    \includegraphics[width= 0.31\textwidth]{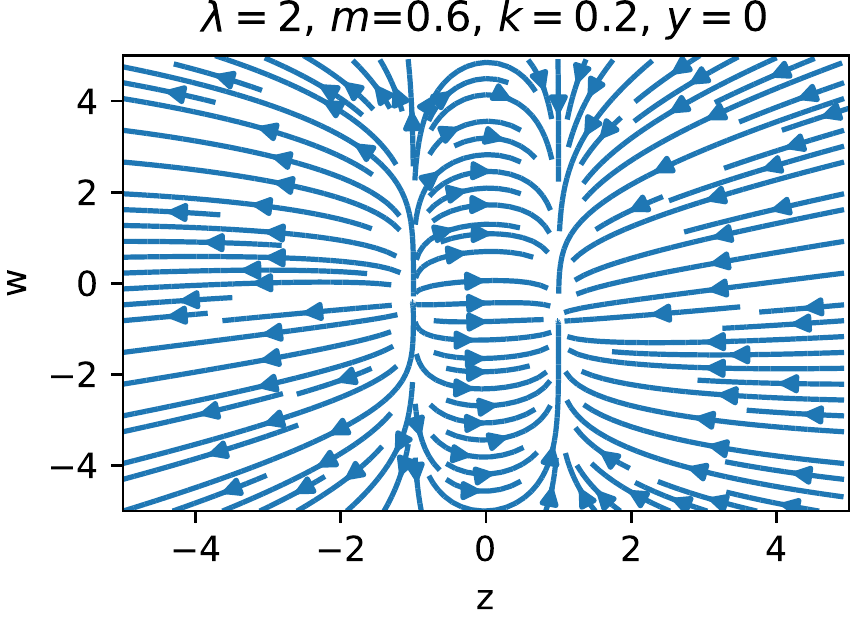}
    \includegraphics[width= 0.32\textwidth]{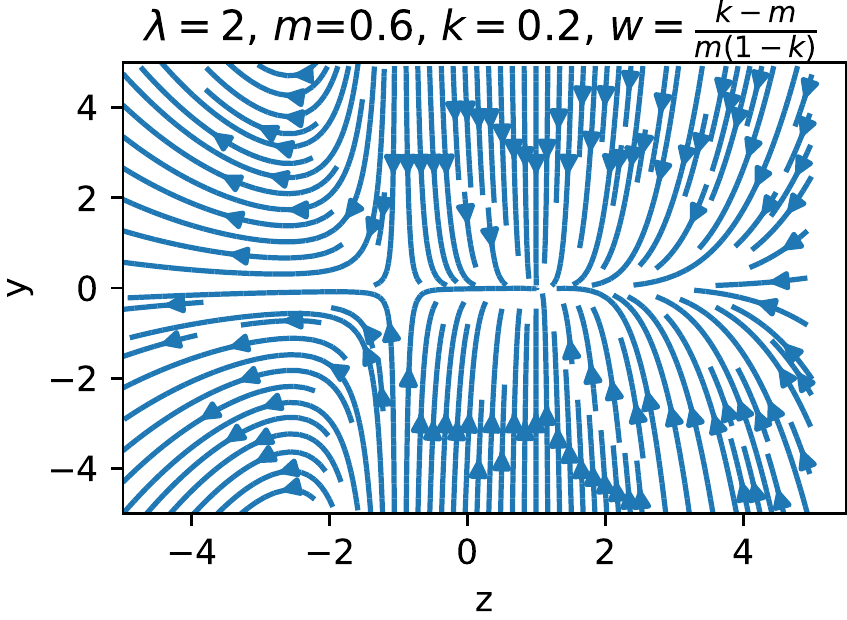}
    \includegraphics[width= 0.31\textwidth]{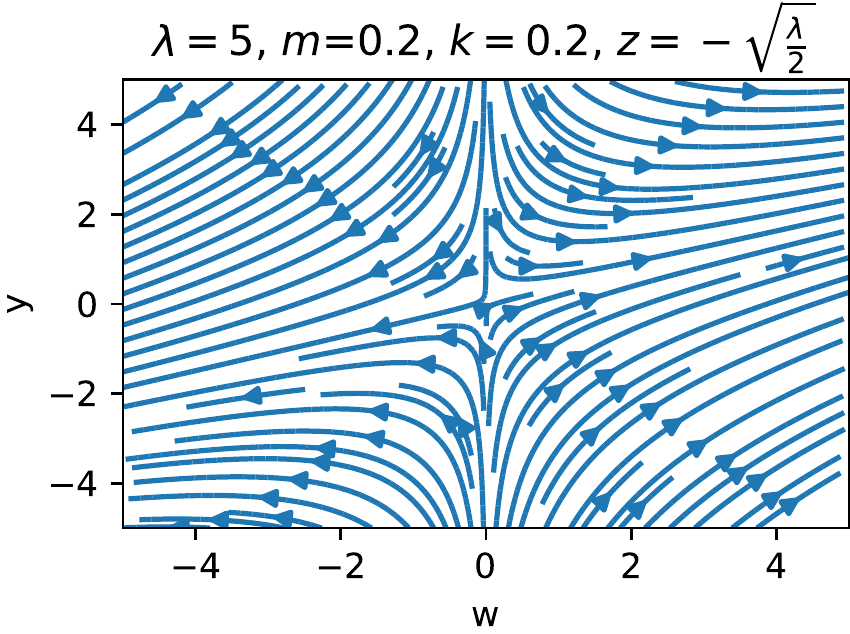}
    \includegraphics[width= 0.32\textwidth]{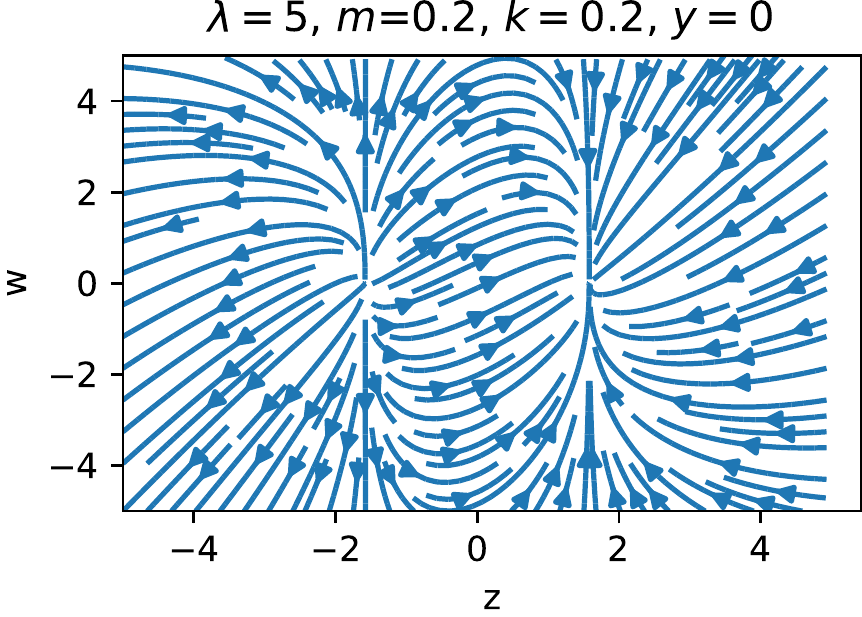}
    \includegraphics[width= 0.31\textwidth]{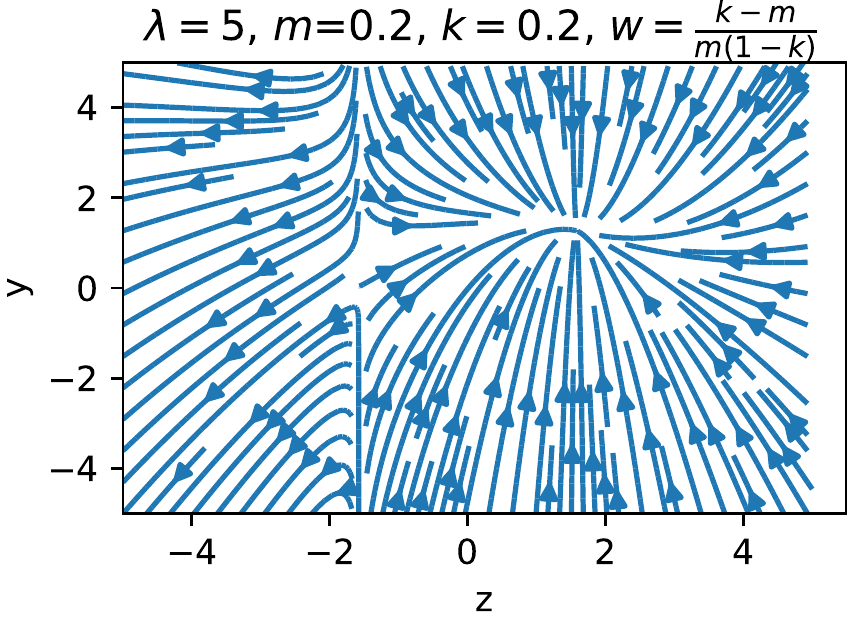}
    \includegraphics[width= 0.31\textwidth]{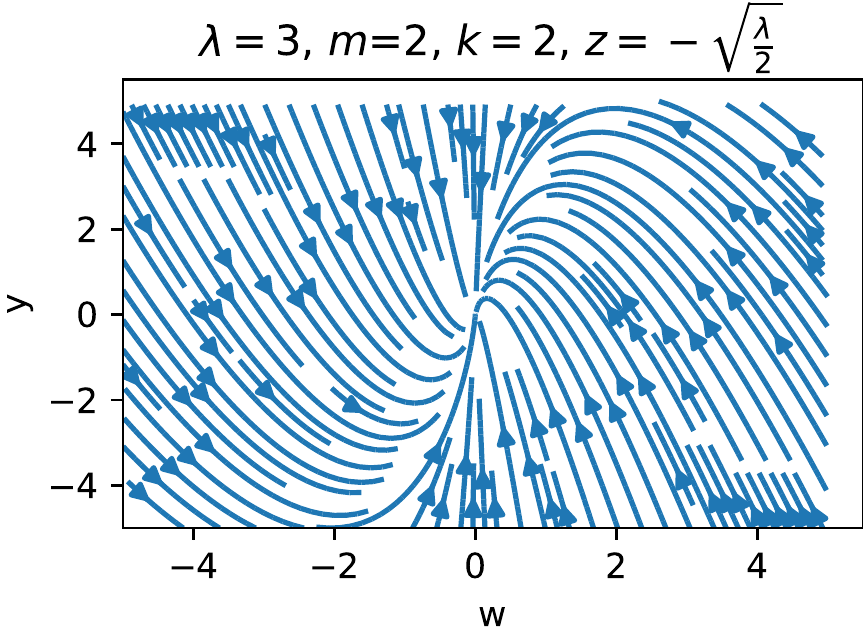}
    \includegraphics[width= 0.31\textwidth]{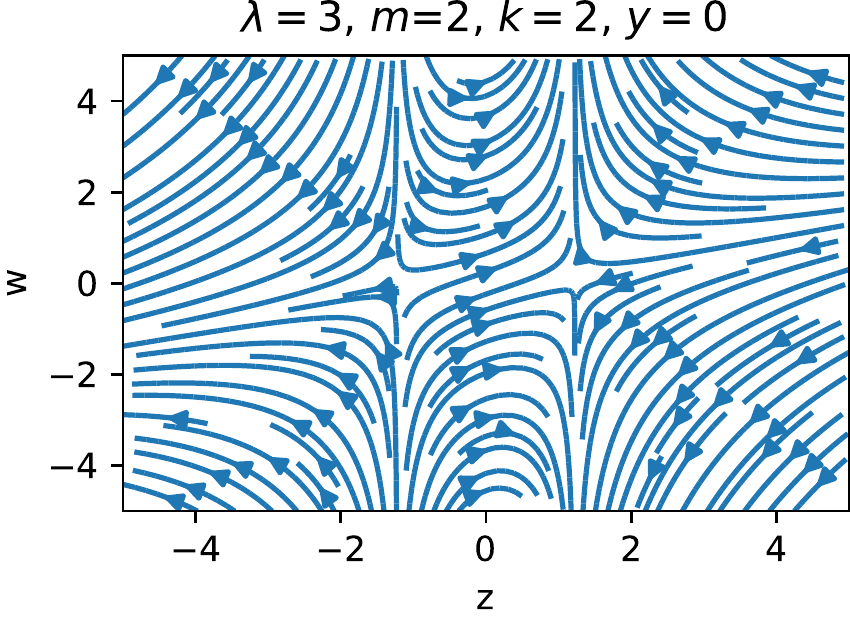}
    \includegraphics[width= 0.31\textwidth]{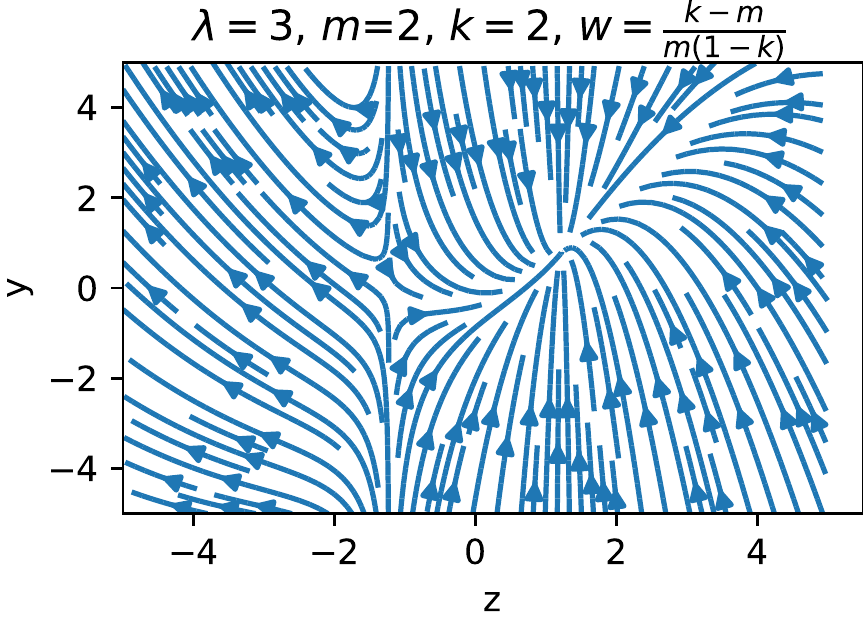}
    \caption{Different views of the phase space of the dynamical system in Eq.(\ref{eq:powerlaw_dynamical}) for $\lambda \neq 0$. The system was reduced to a 2-d surface representing the different perspectives of its constraint with $x = -\frac{k}{m} \Big(\frac{m-1}{k-1}\Big) \frac{1}{3 + z}$. The arrows represent the direction of the velocity field and the trajectories reveal their stability properties as described for this model. 
    }
    \label{fig:power_law_case}
\end{figure*}
\end{widetext}

In Fig.(\ref{fig:power_law_case}) we show the dynamical phase space for this model.

\subsection{Stability analysis for Mixed Power Law Model \label{mixed_model}}
In order to reproduce several important power law scale factors relevant for several cosmological epochs, in Ref.\cite{Bahamonde:2016grb} a form of $f(T,B)$ given by
\begin{equation} \label{eq:mixed_power}
f(T,B) = f_0 B^k T^m\,,
\end{equation}
was presented, where the second and fourth order contributions will now be mixed, and $f_0,k,m$ are arbitrary constants. We can recover GR limit when the index powers vanish, i.e. when $k=0=m$. For this case, the model can be written in terms of the dynamical variables through
\begin{equation}\label{eq:mixed_case}
f_T = -mw\,.
\end{equation}
In comparison to the latter $f(T,B)$ scenarios, this case has the following particularity where
\begin{align}
x = f_B = k f_0 B^{k-1} T^m = \frac{k}{B}f = \frac{k}{6(3H^2 + \dot{H})}f = \frac{f}{6H^2}\frac{k}{3 +\frac{\dot{H}}{H^2}}=-\frac{wk}{3+z}\,,
\end{align}
from which we can notice that $x$ is not an independent variable of the dynamical system. In the same way, when $y = x'$ we obtain directly that $y = y(w,z)$. With these conditions, the autonomous system for this case can be reduced to a 2-d dynamical phase space
\begin{align}
z' = \lambda - 2 z^2\,,  \quad
w' = w\left[\frac{6z(k+m-1) + \lambda k + 2z^2(m-1)}{3+z}\right]\,.
\end{align}
The critical point of the latter system are
\begin{align}
z = \pm \sqrt{\frac{\lambda}{2}}\,, \quad \text{and} \quad w=0\,.
\end{align}
Under these values, the constriction of the system is given by  $\Omega = 1$. Again, we can consider the two roots as follow:

\subsubsection{Critical points}

\begin{enumerate}
\item Positive branch. This case are determinate by the condition $m>0$, with eigenvalues
\begin{align}
\omega_1 = -2 \sqrt{2} \sqrt{\lambda}\,, \\
\omega_2 = (k+m-1)\frac{6\sqrt{\frac{\lambda}{2}} + \lambda}{3 + \sqrt{\frac{\lambda}{2}}}\,.
\end{align}
If $\lambda >0$, we obtain that the eigenvalues are real for any value of $\lambda$, $m$ and $k$. On the other hand, $\omega_2 = 0$ if $k=1-m$, which represent a non-hyperbolic point. We obtain an attractor point if $k < 1-m$, and saddle-like otherwise.

\item Negative branch. The eigenvalues for this case are
\begin{align}
\omega_1 = 2 \sqrt{2} \sqrt{\lambda}\,, \\
\omega_2 = (k+m-1)\frac{-6\sqrt{\frac{\lambda}{2}} + \lambda}{3 - \sqrt{\frac{\lambda}{2}}}\,.
\end{align}
Again, both values are real. 
The critical point is repulsor-like if $k < 1-m$ and $\lambda >0 \land \lambda \neq 18$. We obtain a saddle point if $k > 1-m$ and $\lambda >0 \land \lambda \neq 18$. For $m>k$ or $m<k$ we get a phantom-like EoS ($w<-1$).
\end{enumerate}

\begin{figure}[H]
    \centering
    \includegraphics[width=0.31 \textwidth]{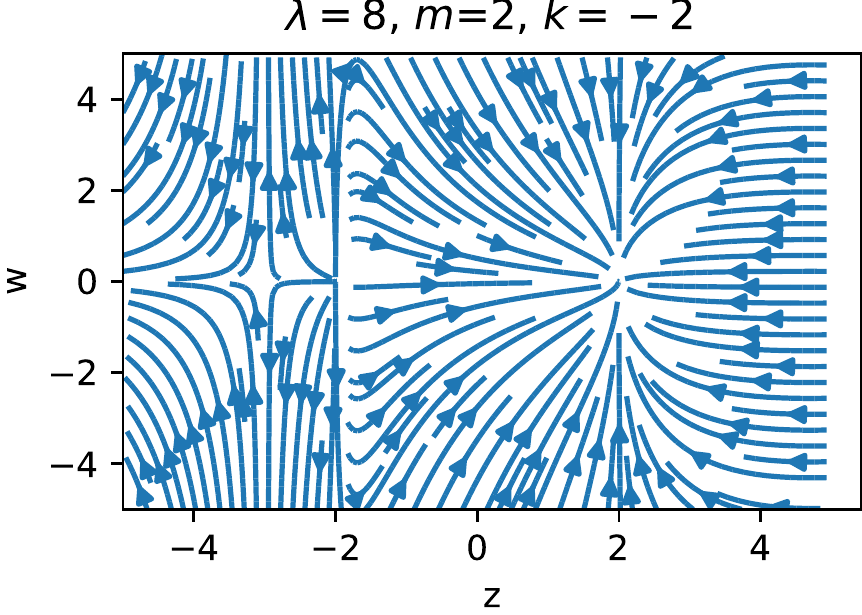}
    \includegraphics[width=0.31 \textwidth]{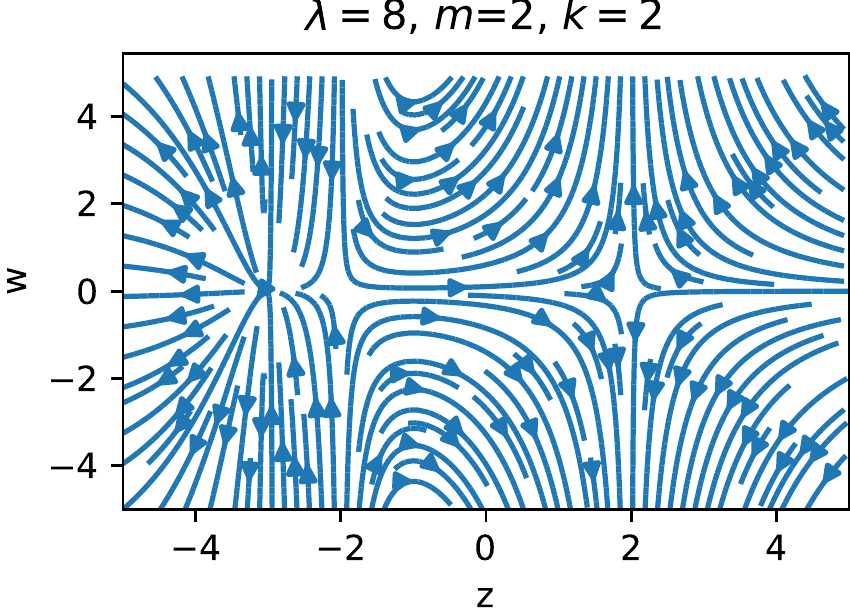}
    \includegraphics[width=0.31 \textwidth]{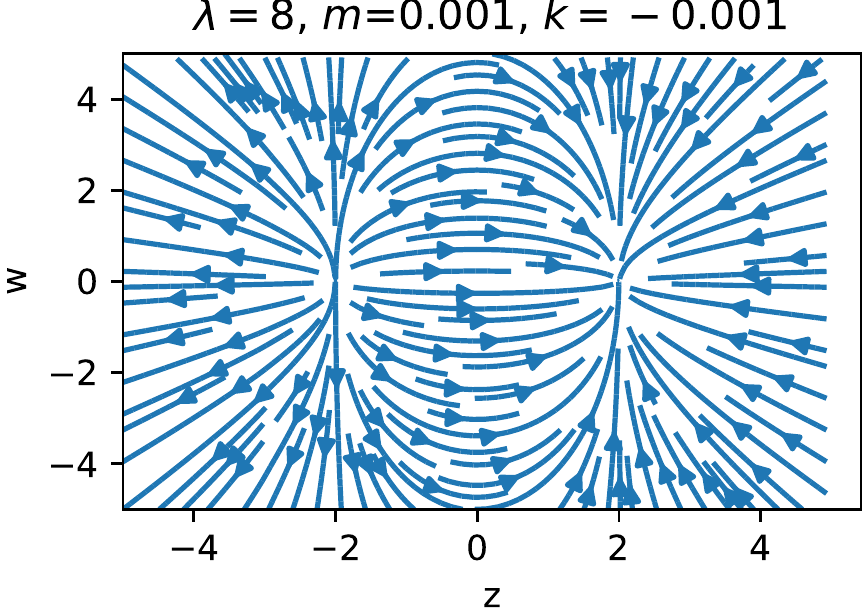}
    \caption{Different views of the phase space of the dynamical system in Eq.(\ref{eq:mixed_case}). The system was reduced to a 2-d surface representing different perspectives. The arrows represent the direction of the velocity field and the trajectories reveal their stability properties as described for this model.}
    \label{mixed_power_law}
\end{figure}

This result reinforces our work in Ref.\cite{Escamilla-Rivera:2019ulu} where we find that for mixed power law models, that the equation of state does cross the phantom line ($\omega=-1$) but preserves the quintessence behaviour up to high redshifts where the model limits to $\Lambda$CDM. For lower redshifts, both $m>k$ and $m<k$ scenarios mimick phantom energy. In fact, these two scenarios correspond to the critical points that we find in the above analysis.

\section{Discussion\label{conc}}
\label{sec:conclusions}

Dynamical systems can reveal a lot of information about the cosmology of theories beyond GR, which may be difficult to study at background or using direct cosmological perturbation theory. In this work, we analysed dynamical systems within the $f(T,B)$ gravity context which was first studied in Refs.\cite{Paliathanasis:2017flf,Paliathanasis:2017efk,Karpathopoulos:2017arc} in a context of scalar field in order to study the implications on inflation solutions. TG offers an avenue to constructing theories which exhibit torsion rather than curvature by exchanging the Levi-Civita connection with its teleparallel connection analog. This produces a wide range of potential cosmological models since TG is naturally lower order and so produces novel manifestations of gravity in addition to those constructed through extensions to GR \cite{Sotiriou:2008rp,Faraoni:2008mf,Capozziello:2011et}.

$f(T,B)$ gravity is an interesting context within which to study dynamical systems since it is one of the rare higher order theories in TG that occurs naturally. Indeed, in section~\ref{f_T_B_dynamic_struc} we outline our strategy in terms of which dynamical variables will produce a suitable dynamical analysis of cosmological systems. $f(T,B)$ gravity acts as a TG generalisation of $f(\lc{R})$ gravity in that the second and fourth order contributions are decoupled from one another through the torsion scalar $T$ and boundary term $B$ with coincidence only for cases where $f(T,B)=f(-T+B)=f(\lc{R})$. The effect of this point also plays out in the dynamical systems analysis where we also must take the dynamical variable defined $\lambda$ in Eq.(\ref{eq:ansatz_lambda}) which is directly analogous to the approach taken in \cite{Odintsov:2018uaw}. Indeed, the analysis where this parameter was probed against possible constant values was studied in \cite{Odintsov:2017tbc}. A feature we explore in this work through the methodology outlined in section~\ref{stabil_method}.

The core results of the work are presented in section~\ref{dynami_ana} where the models (that are cosmological viable at background level) are analysed. We start by probing the general Taylor expansion model in Eq.(\ref{taylor_model}) where the arbitrary function is expanded about the Minkowski values of the scalar arguments up to quadratic order (due to the linear form of $B$ being a boundary term). Here we find the critical points in Eq.(\ref{taylor_crit}) with system eigenvalues in Eq.(\ref{taylor_eigenvalues}). We find that for any nonvanishing constant value of $\lambda$, all the critical values are all saddle points. An interesting feature of this investigation is that the constraints found are consistent with Ref.\cite{Escamilla-Rivera:2019ulu} showing consistency in its confrontation with observations. The evolution for the various parameter combination is shown in Fig.(\ref{fig:taylor_case}).

Afterward, we then study the power law model in section~\ref{power_law_sec} which considers a power law form for both scalar contributors. In this case, the determining factor is the $z$ variable which depends only on derivatives of the Hubble parameter as defined in Eq.(\ref{eq:aut-system}). In this scenario, we find either attractors or saddle points for the positive branch and repellent or saddle points for the negative branch which is shown in Fig.(\ref{fig:power_law_case}).

A similar picture also emerges for the mixed model investigated in section~\ref{mixed_model}. Here, we again find the same variable to be the determining factor in the behaviour of the dynamics of the system. On the other hand, the system turns out to be relatively straightforward to analysis with clear cut results which tally with the general results of the power law model. These are shown in Fig.(\ref{mixed_power_law}). The above results can be linked in a more straightforward manner if we consider directly the form for the Equation of State (EoS). According to the generic EoS reported in  \cite{Escamilla-Rivera:2019ulu} and using our dynamical results, we obtain that for the first two cases (Taylor and Power law model) 
\begin{equation}
\omega_{\text{eff}} = -1 \mp \frac{2}{3} \sqrt{\frac{\lambda}{2}}\,.
\end{equation}
Meanwhile, for the Mixed Power Law model\footnote{We consider an approximate solution to avoid divergence in the critical points obtained for this case.} 
\begin{equation}
\omega_{\text{eff}} \xrightarrow{\text{$\omega \rightarrow 0$}} -1 \mp \frac{2}{3}(k + m)\sqrt{\frac{ \lambda}{2}}\,.   
\end{equation}
Notice that we recover $\Lambda$CDM in the GR limit. In both EoS scenarios, we recover a $\Lambda$CDM model when $\lambda$ vanishes, this can happen when we obtain de Sitter solutions as $H=\text{constant}$. Notice that we can rewrite the $z$-variable from Eq.(\ref{eq:aut-system}) using the definition of the second cosmographic parameter, the deceleration parameter ($q=-\ddot{a}a/\dot{a}^2$) as $\dot{H}=-H^2 (q+1)$, therefore $z=-(q+1)$. In terms of $\lambda$, this latter parameter can be written as  $q = -1 \mp \sqrt{\frac{\lambda}{2}}$, which for $H=\text{constant}$ we obtain that $q=-1$.
Also, we can rewrite the ansatz for $\lambda$ given in Eq.(\ref{eq:ansatz_lambda}) in terms of the third cosmographic parameter, the jerk ($\dddot{a}/aH^3$) as $j = \frac{\ddot{H}}{H^3} -3q -2$, which again in terms of $\lambda$ is $j = \lambda \mp 3\sqrt{\frac{\lambda}{2}} +1 $. Notice that when $\lambda=0$, we recover the standard value $j=1$.

Another important feature of the analysis in this work is the role of couplings between the torsion tensor $T$ and the boundary term $B$. As discussed in the introduction, these represent the second order and fourth order contributions to the field equations respectively, and in a particular combination, forming $f(\lc{R})=f(-T+B)$ gravity. In $f(\lc{R})$ gravity, these couplings do appear but in a very prescribed format. In the present case we allow for more novel models to develop. In particular in cases 1 and 3 of section~\ref{dynami_ana}, the coupling term in the Lagrangian plays an impactful role in the dynamics that ensue. This is an interesting property that should be investigated further.

From the results obtained with this proposal, we notice that it will be interesting to study the behaviour of other $f(T,B)$ gravity models, which together with their confrontation with observational data, may open an avenue for producing other viable cosmological scenarios. Furthermore, the role of a varying $\lambda$ is also an important future work which may better expose the dynamical behaviour of $f(T,B)$ gravity, since from the Taylor and Power Law cases we will require a non-autonomous system with $\lambda\neq \text{const}$. Another important higher-order extension to TG is $f(T,T_G)$ which may also have interesting properties. This study will be reported elsewhere. 


\begin{acknowledgments}

CE-R acknowledges the Royal Astronomical Society as FRAS 10147 and PAPIIT Project IA100220. CE-R and JLS would like to acknowledge networking support by the COST Action CA18108. JLS would also like to acknowledge funding support from Cosmology@MALTA which is supported by the University of Malta.

\end{acknowledgments}

\bibliography{librero0}

\begin{thebibliography}{66}
\expandafter\ifx\csname natexlab\endcsname\relax\def\natexlab#1{#1}\fi
\expandafter\ifx\csname bibnamefont\endcsname\relax
  \def\bibnamefont#1{#1}\fi
\expandafter\ifx\csname bibfnamefont\endcsname\relax
  \def\bibfnamefont#1{#1}\fi
\expandafter\ifx\csname citenamefont\endcsname\relax
  \def\citenamefont#1{#1}\fi
\expandafter\ifx\csname url\endcsname\relax
  \def\url#1{\texttt{#1}}\fi
\expandafter\ifx\csname urlprefix\endcsname\relax\def\urlprefix{URL }\fi
\providecommand{\bibinfo}[2]{#2}
\providecommand{\eprint}[2][]{\url{#2}}

\bibitem[{\citenamefont{Misner et~al.}(1973)\citenamefont{Misner, Thorne, and
  Wheeler}}]{misner1973gravitation}
\bibinfo{author}{\bibfnamefont{C.}~\bibnamefont{Misner}},
  \bibinfo{author}{\bibfnamefont{K.}~\bibnamefont{Thorne}}, \bibnamefont{and}
  \bibinfo{author}{\bibfnamefont{J.}~\bibnamefont{Wheeler}}
  (\bibinfo{year}{1973}),
  \urlprefix\url{https://books.google.com.mt/books?id=w4Gigq3tY1kC}.

\bibitem[{\citenamefont{Clifton et~al.}(2012)\citenamefont{Clifton, Ferreira,
  Padilla, and Skordis}}]{Clifton:2011jh}
\bibinfo{author}{\bibfnamefont{T.}~\bibnamefont{Clifton}},
  \bibinfo{author}{\bibfnamefont{P.~G.} \bibnamefont{Ferreira}},
  \bibinfo{author}{\bibfnamefont{A.}~\bibnamefont{Padilla}}, \bibnamefont{and}
  \bibinfo{author}{\bibfnamefont{C.}~\bibnamefont{Skordis}},
  \bibinfo{journal}{Phys. Rept.} \textbf{\bibinfo{volume}{513}},
  \bibinfo{pages}{1} (\bibinfo{year}{2012}), \eprint{1106.2476}.

\bibitem[{\citenamefont{Baudis}(2016)}]{Baudis:2016qwx}
\bibinfo{author}{\bibfnamefont{L.}~\bibnamefont{Baudis}}, \bibinfo{journal}{J.
  Phys.} \textbf{\bibinfo{volume}{G43}}, \bibinfo{pages}{044001}
  (\bibinfo{year}{2016}).

\bibitem[{\citenamefont{Bertone et~al.}(2005)\citenamefont{Bertone, Hooper, and
  Silk}}]{Bertone:2004pz}
\bibinfo{author}{\bibfnamefont{G.}~\bibnamefont{Bertone}},
  \bibinfo{author}{\bibfnamefont{D.}~\bibnamefont{Hooper}}, \bibnamefont{and}
  \bibinfo{author}{\bibfnamefont{J.}~\bibnamefont{Silk}},
  \bibinfo{journal}{Phys. Rept.} \textbf{\bibinfo{volume}{405}},
  \bibinfo{pages}{279} (\bibinfo{year}{2005}), \eprint{hep-ph/0404175}.

\bibitem[{\citenamefont{Peebles and Ratra}(2003)}]{Peebles:2002gy}
\bibinfo{author}{\bibfnamefont{P.~J.~E.} \bibnamefont{Peebles}}
  \bibnamefont{and} \bibinfo{author}{\bibfnamefont{B.}~\bibnamefont{Ratra}},
  \bibinfo{journal}{Rev. Mod. Phys.} \textbf{\bibinfo{volume}{75}},
  \bibinfo{pages}{559} (\bibinfo{year}{2003}), \bibinfo{note}{[,592(2002)]},
  \eprint{astro-ph/0207347}.

\bibitem[{\citenamefont{Copeland et~al.}(2006)\citenamefont{Copeland, Sami, and
  Tsujikawa}}]{Copeland:2006wr}
\bibinfo{author}{\bibfnamefont{E.~J.} \bibnamefont{Copeland}},
  \bibinfo{author}{\bibfnamefont{M.}~\bibnamefont{Sami}}, \bibnamefont{and}
  \bibinfo{author}{\bibfnamefont{S.}~\bibnamefont{Tsujikawa}},
  \bibinfo{journal}{Int. J. Mod. Phys.} \textbf{\bibinfo{volume}{D15}},
  \bibinfo{pages}{1753} (\bibinfo{year}{2006}), \eprint{hep-th/0603057}.

\bibitem[{\citenamefont{Riess et~al.}(1998)}]{Riess:1998cb}
\bibinfo{author}{\bibfnamefont{A.~G.} \bibnamefont{Riess}} \bibnamefont{et~al.}
  (\bibinfo{collaboration}{Supernova Search Team}),
  \bibinfo{journal}{Astron.J.} \textbf{\bibinfo{volume}{116}},
  \bibinfo{pages}{1009} (\bibinfo{year}{1998}), \eprint{astro-ph/9805201}.

\bibitem[{\citenamefont{Perlmutter et~al.}(1999)}]{Perlmutter:1998np}
\bibinfo{author}{\bibfnamefont{S.}~\bibnamefont{Perlmutter}}
  \bibnamefont{et~al.} (\bibinfo{collaboration}{Supernova Cosmology Project}),
  \bibinfo{journal}{Astrophys.J.} \textbf{\bibinfo{volume}{517}},
  \bibinfo{pages}{565} (\bibinfo{year}{1999}), \eprint{astro-ph/9812133}.

\bibitem[{\citenamefont{Weinberg}(1989)}]{RevModPhys.61.1}
\bibinfo{author}{\bibfnamefont{S.}~\bibnamefont{Weinberg}},
  \bibinfo{journal}{Rev. Mod. Phys.} \textbf{\bibinfo{volume}{61}},
  \bibinfo{pages}{1} (\bibinfo{year}{1989}),
  \urlprefix\url{https://link.aps.org/doi/10.1103/RevModPhys.61.1}.

\bibitem[{\citenamefont{Gaitskell}(2004)}]{Gaitskell:2004gd}
\bibinfo{author}{\bibfnamefont{R.}~\bibnamefont{Gaitskell}},
  \bibinfo{journal}{Ann.\ Rev.\ Nucl.\ Part.\ Sci.}
  \textbf{\bibinfo{volume}{54}}, \bibinfo{pages}{315} (\bibinfo{year}{2004}).

\bibitem[{\citenamefont{Riess et~al.}(2019)\citenamefont{Riess, Casertano,
  Yuan, Macri, and Scolnic}}]{Riess:2019cxk}
\bibinfo{author}{\bibfnamefont{A.~G.} \bibnamefont{Riess}},
  \bibinfo{author}{\bibfnamefont{S.}~\bibnamefont{Casertano}},
  \bibinfo{author}{\bibfnamefont{W.}~\bibnamefont{Yuan}},
  \bibinfo{author}{\bibfnamefont{L.~M.} \bibnamefont{Macri}}, \bibnamefont{and}
  \bibinfo{author}{\bibfnamefont{D.}~\bibnamefont{Scolnic}},
  \bibinfo{journal}{Astrophys. J.} \textbf{\bibinfo{volume}{876}},
  \bibinfo{pages}{85} (\bibinfo{year}{2019}), \eprint{1903.07603}.

\bibitem[{\citenamefont{Wong et~al.}(2019)}]{Wong:2019kwg}
\bibinfo{author}{\bibfnamefont{K.~C.} \bibnamefont{Wong}} \bibnamefont{et~al.}
  (\bibinfo{year}{2019}), \eprint{1907.04869}.

\bibitem[{\citenamefont{Aghanim et~al.}(2018)}]{Aghanim:2018eyx}
\bibinfo{author}{\bibfnamefont{N.}~\bibnamefont{Aghanim}} \bibnamefont{et~al.}
  (\bibinfo{collaboration}{Planck}) (\bibinfo{year}{2018}),
  \eprint{1807.06209}.

\bibitem[{\citenamefont{Ade et~al.}(2016)}]{Ade:2015xua}
\bibinfo{author}{\bibfnamefont{P.}~\bibnamefont{Ade}} \bibnamefont{et~al.}
  (\bibinfo{collaboration}{Planck}), \bibinfo{journal}{Astron.Astrophys.}
  \textbf{\bibinfo{volume}{594}}, \bibinfo{pages}{A13} (\bibinfo{year}{2016}),
  \eprint{1502.01589}.

\bibitem[{\citenamefont{Gómez-Valent and Amendola}()}]{Gomez-Valent:2019lny}
\bibinfo{author}{\bibfnamefont{A.}~\bibnamefont{Gómez-Valent}}
  \bibnamefont{and} \bibinfo{author}{\bibfnamefont{L.}~\bibnamefont{Amendola}}
  (????).

\bibitem[{\citenamefont{Graef et~al.}(2019)\citenamefont{Graef, Benetti, and
  Alcaniz}}]{Graef:2018fzu}
\bibinfo{author}{\bibfnamefont{L.~L.} \bibnamefont{Graef}},
  \bibinfo{author}{\bibfnamefont{M.}~\bibnamefont{Benetti}}, \bibnamefont{and}
  \bibinfo{author}{\bibfnamefont{J.~S.} \bibnamefont{Alcaniz}},
  \bibinfo{journal}{Phys.Rev.D} \textbf{\bibinfo{volume}{99}},
  \bibinfo{pages}{043519} (\bibinfo{year}{2019}), \eprint{1809.04501}.

\bibitem[{\citenamefont{Abbott et~al.}(2017)}]{Abbott:2017xzu}
\bibinfo{author}{\bibfnamefont{B.}~\bibnamefont{Abbott}} \bibnamefont{et~al.}
  (\bibinfo{collaboration}{LIGO Scientific, Virgo, 1M2H, Dark Energy Camera
  GW-E, DES, DLT40, Las Cumbres Observatory, VINROUGE, MASTER}),
  \bibinfo{journal}{Nature} \textbf{\bibinfo{volume}{551}}, \bibinfo{pages}{85}
  (\bibinfo{year}{2017}), \eprint{1710.05835}.

\bibitem[{\citenamefont{Baker et~al.}(2019)}]{Baker:2019nia}
\bibinfo{author}{\bibfnamefont{J.}~\bibnamefont{Baker}} \bibnamefont{et~al.}
  (\bibinfo{year}{2019}), \eprint{1907.06482}.

\bibitem[{\citenamefont{{Amaro-Seoane} et~al.}(2017)}]{2017arXiv170200786A}
\bibinfo{author}{\bibfnamefont{P.}~\bibnamefont{{Amaro-Seoane}}}
  \bibnamefont{et~al.}, \bibinfo{journal}{arXiv e-prints}
  \bibinfo{eid}{arXiv:1702.00786} (\bibinfo{year}{2017}), \eprint{1702.00786}.

\bibitem[{\citenamefont{Bahamonde
  et~al.}(2018{\natexlab{a}})\citenamefont{Bahamonde, Böhmer, Carloni,
  Copeland, Fang, and Tamanini}}]{Bahamonde:2017ize}
\bibinfo{author}{\bibfnamefont{S.}~\bibnamefont{Bahamonde}},
  \bibinfo{author}{\bibfnamefont{C.~G.} \bibnamefont{Böhmer}},
  \bibinfo{author}{\bibfnamefont{S.}~\bibnamefont{Carloni}},
  \bibinfo{author}{\bibfnamefont{E.~J.} \bibnamefont{Copeland}},
  \bibinfo{author}{\bibfnamefont{W.}~\bibnamefont{Fang}}, \bibnamefont{and}
  \bibinfo{author}{\bibfnamefont{N.}~\bibnamefont{Tamanini}},
  \bibinfo{journal}{Phys. Rept.} \textbf{\bibinfo{volume}{775-777}},
  \bibinfo{pages}{1} (\bibinfo{year}{2018}{\natexlab{a}}), \eprint{1712.03107}.

\bibitem[{\citenamefont{Capozziello and
  De~Laurentis}(2011)}]{Capozziello:2011et}
\bibinfo{author}{\bibfnamefont{S.}~\bibnamefont{Capozziello}} \bibnamefont{and}
  \bibinfo{author}{\bibfnamefont{M.}~\bibnamefont{De~Laurentis}},
  \bibinfo{journal}{Phys. Rept.} \textbf{\bibinfo{volume}{509}},
  \bibinfo{pages}{167} (\bibinfo{year}{2011}), \eprint{1108.6266}.

\bibitem[{\citenamefont{Sotiriou and Faraoni}(2010)}]{Sotiriou:2008rp}
\bibinfo{author}{\bibfnamefont{T.~P.} \bibnamefont{Sotiriou}} \bibnamefont{and}
  \bibinfo{author}{\bibfnamefont{V.}~\bibnamefont{Faraoni}},
  \bibinfo{journal}{Rev. Mod. Phys.} \textbf{\bibinfo{volume}{82}},
  \bibinfo{pages}{451} (\bibinfo{year}{2010}), \eprint{0805.1726}.

\bibitem[{\citenamefont{Faraoni}(2008)}]{Faraoni:2008mf}
\bibinfo{author}{\bibfnamefont{V.}~\bibnamefont{Faraoni}}
  (\bibinfo{year}{2008}), \eprint{0810.2602}.

\bibitem[{\citenamefont{Nakahara}(2003)}]{nakahara2003geometry}
\bibinfo{author}{\bibfnamefont{M.}~\bibnamefont{Nakahara}}
  (\bibinfo{year}{2003}),
  \urlprefix\url{https://books.google.com.mt/books?id=cH-XQB0Ex5wC}.

\bibitem[{\citenamefont{Aldrovandi and Pereira}(2013)}]{Aldrovandi:2013wha}
\bibinfo{author}{\bibfnamefont{R.}~\bibnamefont{Aldrovandi}} \bibnamefont{and}
  \bibinfo{author}{\bibfnamefont{J.~G.} \bibnamefont{Pereira}},
  \bibinfo{journal}{Fundam. Theor. Phys.} \textbf{\bibinfo{volume}{173}}
  (\bibinfo{year}{2013}).

\bibitem[{\citenamefont{Cai et~al.}(2016)\citenamefont{Cai, Capozziello,
  De~Laurentis, and Saridakis}}]{Cai:2015emx}
\bibinfo{author}{\bibfnamefont{Y.-F.} \bibnamefont{Cai}},
  \bibinfo{author}{\bibfnamefont{S.}~\bibnamefont{Capozziello}},
  \bibinfo{author}{\bibfnamefont{M.}~\bibnamefont{De~Laurentis}},
  \bibnamefont{and} \bibinfo{author}{\bibfnamefont{E.~N.}
  \bibnamefont{Saridakis}}, \bibinfo{journal}{Rept. Prog. Phys.}
  \textbf{\bibinfo{volume}{79}}, \bibinfo{pages}{106901}
  (\bibinfo{year}{2016}), \eprint{1511.07586}.

\bibitem[{\citenamefont{Kr\v{s}\v{s}\'{a}k
  et~al.}(2019)\citenamefont{Kr\v{s}\v{s}\'{a}k, van~den Hoogen, Pereira,
  Böhmer, and Coley}}]{Krssak:2018ywd}
\bibinfo{author}{\bibfnamefont{M.}~\bibnamefont{Kr\v{s}\v{s}\'{a}k}},
  \bibinfo{author}{\bibfnamefont{R.~J.} \bibnamefont{van~den Hoogen}},
  \bibinfo{author}{\bibfnamefont{J.~G.} \bibnamefont{Pereira}},
  \bibinfo{author}{\bibfnamefont{C.~G.} \bibnamefont{Böhmer}},
  \bibnamefont{and} \bibinfo{author}{\bibfnamefont{A.~A.} \bibnamefont{Coley}},
  \bibinfo{journal}{Class. Quant. Grav.} \textbf{\bibinfo{volume}{36}},
  \bibinfo{pages}{183001} (\bibinfo{year}{2019}), \eprint{1810.12932}.

\bibitem[{\citenamefont{Weitzenb\"{o}ock}(1923)}]{Weitzenbock1923}
\bibinfo{author}{\bibfnamefont{R.}~\bibnamefont{Weitzenb\"{o}ock}},
  \emph{\bibinfo{title}{Invariantentheorie}} (\bibinfo{publisher}{Noordhoff,
  Gronningen}, \bibinfo{year}{1923}).

\bibitem[{\citenamefont{Lovelock}(1971)}]{Lovelock:1971yv}
\bibinfo{author}{\bibfnamefont{D.}~\bibnamefont{Lovelock}},
  \bibinfo{journal}{J. Math. Phys.} \textbf{\bibinfo{volume}{12}},
  \bibinfo{pages}{498} (\bibinfo{year}{1971}).

\bibitem[{\citenamefont{Gonzalez and Vasquez}(2015)}]{Gonzalez:2015sha}
\bibinfo{author}{\bibfnamefont{P.~A.} \bibnamefont{Gonzalez}} \bibnamefont{and}
  \bibinfo{author}{\bibfnamefont{Y.}~\bibnamefont{Vasquez}},
  \bibinfo{journal}{Phys. Rev.} \textbf{\bibinfo{volume}{D92}},
  \bibinfo{pages}{124023} (\bibinfo{year}{2015}), \eprint{1508.01174}.

\bibitem[{\citenamefont{Bahamonde
  et~al.}(2019{\natexlab{a}})\citenamefont{Bahamonde, Dialektopoulos, and
  Levi~Said}}]{Bahamonde:2019shr}
\bibinfo{author}{\bibfnamefont{S.}~\bibnamefont{Bahamonde}},
  \bibinfo{author}{\bibfnamefont{K.~F.} \bibnamefont{Dialektopoulos}},
  \bibnamefont{and}
  \bibinfo{author}{\bibfnamefont{J.}~\bibnamefont{Levi~Said}},
  \bibinfo{journal}{Phys. Rev.} \textbf{\bibinfo{volume}{D100}},
  \bibinfo{pages}{064018} (\bibinfo{year}{2019}{\natexlab{a}}),
  \eprint{1904.10791}.

\bibitem[{\citenamefont{Blixt et~al.}(2019{\natexlab{a}})\citenamefont{Blixt,
  Hohmann, and Pfeifer}}]{Blixt:2018znp}
\bibinfo{author}{\bibfnamefont{D.}~\bibnamefont{Blixt}},
  \bibinfo{author}{\bibfnamefont{M.}~\bibnamefont{Hohmann}}, \bibnamefont{and}
  \bibinfo{author}{\bibfnamefont{C.}~\bibnamefont{Pfeifer}},
  \bibinfo{journal}{Phys. Rev.} \textbf{\bibinfo{volume}{D99}},
  \bibinfo{pages}{084025} (\bibinfo{year}{2019}{\natexlab{a}}),
  \eprint{1811.11137}.

\bibitem[{\citenamefont{Blixt et~al.}(2019{\natexlab{b}})\citenamefont{Blixt,
  Hohmann, and Pfeifer}}]{Blixt:2019mkt}
\bibinfo{author}{\bibfnamefont{D.}~\bibnamefont{Blixt}},
  \bibinfo{author}{\bibfnamefont{M.}~\bibnamefont{Hohmann}}, \bibnamefont{and}
  \bibinfo{author}{\bibfnamefont{C.}~\bibnamefont{Pfeifer}},
  \bibinfo{journal}{Universe} \textbf{\bibinfo{volume}{5}},
  \bibinfo{pages}{143} (\bibinfo{year}{2019}{\natexlab{b}}),
  \eprint{1905.01048}.

\bibitem[{\citenamefont{Ferraro and Fiorini}(2007)}]{Ferraro:2006jd}
\bibinfo{author}{\bibfnamefont{R.}~\bibnamefont{Ferraro}} \bibnamefont{and}
  \bibinfo{author}{\bibfnamefont{F.}~\bibnamefont{Fiorini}},
  \bibinfo{journal}{Phys. Rev.} \textbf{\bibinfo{volume}{D75}},
  \bibinfo{pages}{084031} (\bibinfo{year}{2007}), \eprint{gr-qc/0610067}.

\bibitem[{\citenamefont{Ferraro and Fiorini}(2008)}]{Ferraro:2008ey}
\bibinfo{author}{\bibfnamefont{R.}~\bibnamefont{Ferraro}} \bibnamefont{and}
  \bibinfo{author}{\bibfnamefont{F.}~\bibnamefont{Fiorini}},
  \bibinfo{journal}{Phys. Rev.} \textbf{\bibinfo{volume}{D78}},
  \bibinfo{pages}{124019} (\bibinfo{year}{2008}), \eprint{0812.1981}.

\bibitem[{\citenamefont{Bengochea and Ferraro}(2009)}]{Bengochea:2008gz}
\bibinfo{author}{\bibfnamefont{G.~R.} \bibnamefont{Bengochea}}
  \bibnamefont{and} \bibinfo{author}{\bibfnamefont{R.}~\bibnamefont{Ferraro}},
  \bibinfo{journal}{Phys. Rev.} \textbf{\bibinfo{volume}{D79}},
  \bibinfo{pages}{124019} (\bibinfo{year}{2009}), \eprint{0812.1205}.

\bibitem[{\citenamefont{Linder}(2010)}]{Linder:2010py}
\bibinfo{author}{\bibfnamefont{E.~V.} \bibnamefont{Linder}},
  \bibinfo{journal}{Phys. Rev.} \textbf{\bibinfo{volume}{D81}},
  \bibinfo{pages}{127301} (\bibinfo{year}{2010}), \bibinfo{note}{[Erratum:
  Phys. Rev.D82,109902(2010)]}, \eprint{1005.3039}.

\bibitem[{\citenamefont{Chen et~al.}(2011)\citenamefont{Chen, Dent, Dutta, and
  Saridakis}}]{Chen:2010va}
\bibinfo{author}{\bibfnamefont{S.-H.} \bibnamefont{Chen}},
  \bibinfo{author}{\bibfnamefont{J.~B.} \bibnamefont{Dent}},
  \bibinfo{author}{\bibfnamefont{S.}~\bibnamefont{Dutta}}, \bibnamefont{and}
  \bibinfo{author}{\bibfnamefont{E.~N.} \bibnamefont{Saridakis}},
  \bibinfo{journal}{Phys. Rev.} \textbf{\bibinfo{volume}{D83}},
  \bibinfo{pages}{023508} (\bibinfo{year}{2011}), \eprint{1008.1250}.

\bibitem[{\citenamefont{Bahamonde
  et~al.}(2019{\natexlab{b}})\citenamefont{Bahamonde, Flathmann, and
  Pfeifer}}]{Bahamonde:2019zea}
\bibinfo{author}{\bibfnamefont{S.}~\bibnamefont{Bahamonde}},
  \bibinfo{author}{\bibfnamefont{K.}~\bibnamefont{Flathmann}},
  \bibnamefont{and} \bibinfo{author}{\bibfnamefont{C.}~\bibnamefont{Pfeifer}},
  \bibinfo{journal}{Phys. Rev. D} \textbf{\bibinfo{volume}{100}},
  \bibinfo{pages}{084064} (\bibinfo{year}{2019}{\natexlab{b}}),
  \eprint{1907.10858}.

\bibitem[{\citenamefont{Nesseris et~al.}(2013)\citenamefont{Nesseris,
  Basilakos, Saridakis, and Perivolaropoulos}}]{Nesseris:2013jea}
\bibinfo{author}{\bibfnamefont{S.}~\bibnamefont{Nesseris}},
  \bibinfo{author}{\bibfnamefont{S.}~\bibnamefont{Basilakos}},
  \bibinfo{author}{\bibfnamefont{E.~N.} \bibnamefont{Saridakis}},
  \bibnamefont{and}
  \bibinfo{author}{\bibfnamefont{L.}~\bibnamefont{Perivolaropoulos}},
  \bibinfo{journal}{Phys. Rev.} \textbf{\bibinfo{volume}{D88}},
  \bibinfo{pages}{103010} (\bibinfo{year}{2013}), \eprint{1308.6142}.

\bibitem[{\citenamefont{Farrugia and Said}(2016)}]{Farrugia:2016qqe}
\bibinfo{author}{\bibfnamefont{G.}~\bibnamefont{Farrugia}} \bibnamefont{and}
  \bibinfo{author}{\bibfnamefont{J.~L.} \bibnamefont{Said}},
  \bibinfo{journal}{Phys. Rev.} \textbf{\bibinfo{volume}{D94}},
  \bibinfo{pages}{124054} (\bibinfo{year}{2016}), \eprint{1701.00134}.

\bibitem[{\citenamefont{Finch and Said}(2018)}]{Finch:2018gkh}
\bibinfo{author}{\bibfnamefont{A.}~\bibnamefont{Finch}} \bibnamefont{and}
  \bibinfo{author}{\bibfnamefont{J.~L.} \bibnamefont{Said}},
  \bibinfo{journal}{Eur.Phys.J.C} \textbf{\bibinfo{volume}{78}},
  \bibinfo{pages}{560} (\bibinfo{year}{2018}), \eprint{1806.09677}.

\bibitem[{\citenamefont{Farrugia et~al.}(2016)\citenamefont{Farrugia, Said, and
  Ruggiero}}]{Farrugia:2016xcw}
\bibinfo{author}{\bibfnamefont{G.}~\bibnamefont{Farrugia}},
  \bibinfo{author}{\bibfnamefont{J.~L.} \bibnamefont{Said}}, \bibnamefont{and}
  \bibinfo{author}{\bibfnamefont{M.~L.} \bibnamefont{Ruggiero}},
  \bibinfo{journal}{Phys. Rev.} \textbf{\bibinfo{volume}{D93}},
  \bibinfo{pages}{104034} (\bibinfo{year}{2016}), \eprint{1605.07614}.

\bibitem[{\citenamefont{Iorio and Saridakis}(2012)}]{Iorio:2012cm}
\bibinfo{author}{\bibfnamefont{L.}~\bibnamefont{Iorio}} \bibnamefont{and}
  \bibinfo{author}{\bibfnamefont{E.~N.} \bibnamefont{Saridakis}},
  \bibinfo{journal}{Mon. Not. Roy. Astron. Soc.}
  \textbf{\bibinfo{volume}{427}}, \bibinfo{pages}{1555} (\bibinfo{year}{2012}),
  \eprint{1203.5781}.

\bibitem[{\citenamefont{Ruggiero and Radicella}(2015)}]{Ruggiero:2015oka}
\bibinfo{author}{\bibfnamefont{M.~L.} \bibnamefont{Ruggiero}} \bibnamefont{and}
  \bibinfo{author}{\bibfnamefont{N.}~\bibnamefont{Radicella}},
  \bibinfo{journal}{Phys. Rev.} \textbf{\bibinfo{volume}{D91}},
  \bibinfo{pages}{104014} (\bibinfo{year}{2015}), \eprint{1501.02198}.

\bibitem[{\citenamefont{Deng}(2018)}]{Deng:2018ncg}
\bibinfo{author}{\bibfnamefont{X.-M.} \bibnamefont{Deng}},
  \bibinfo{journal}{Class.Quant.Grav.} \textbf{\bibinfo{volume}{35}},
  \bibinfo{pages}{175013} (\bibinfo{year}{2018}).

\bibitem[{\citenamefont{Paliathanasis}(2017{\natexlab{a}})}]{Bahamonde:2015zma}
\bibinfo{author}{\bibfnamefont{A.}~\bibnamefont{Paliathanasis}},
  \bibinfo{journal}{JCAP} \textbf{\bibinfo{volume}{1708}}, \bibinfo{pages}{027}
  (\bibinfo{year}{2017}{\natexlab{a}}).

\bibitem[{\citenamefont{Capozziello et~al.}(2018)\citenamefont{Capozziello,
  Capriolo, and Transirico}}]{Capozziello:2018qcp}
\bibinfo{author}{\bibfnamefont{S.}~\bibnamefont{Capozziello}},
  \bibinfo{author}{\bibfnamefont{M.}~\bibnamefont{Capriolo}}, \bibnamefont{and}
  \bibinfo{author}{\bibfnamefont{M.}~\bibnamefont{Transirico}},
  \bibinfo{journal}{Int. J. Geom. Meth. Mod. Phys.}
  \textbf{\bibinfo{volume}{15}}, \bibinfo{pages}{1850164}
  (\bibinfo{year}{2018}), \eprint{1804.08530}.

\bibitem[{\citenamefont{Bahamonde and Capozziello}(2017)}]{Bahamonde:2016grb}
\bibinfo{author}{\bibfnamefont{S.}~\bibnamefont{Bahamonde}} \bibnamefont{and}
  \bibinfo{author}{\bibfnamefont{S.}~\bibnamefont{Capozziello}},
  \bibinfo{journal}{Eur. Phys. J.} \textbf{\bibinfo{volume}{C77}},
  \bibinfo{pages}{107} (\bibinfo{year}{2017}), \eprint{1612.01299}.

\bibitem[{\citenamefont{Farrugia et~al.}(2018)\citenamefont{Farrugia,
  Levi~Said, Gakis, and Saridakis}}]{Farrugia:2018gyz}
\bibinfo{author}{\bibfnamefont{G.}~\bibnamefont{Farrugia}},
  \bibinfo{author}{\bibfnamefont{J.}~\bibnamefont{Levi~Said}},
  \bibinfo{author}{\bibfnamefont{V.}~\bibnamefont{Gakis}}, \bibnamefont{and}
  \bibinfo{author}{\bibfnamefont{E.~N.} \bibnamefont{Saridakis}},
  \bibinfo{journal}{Phys. Rev.} \textbf{\bibinfo{volume}{D97}},
  \bibinfo{pages}{124064} (\bibinfo{year}{2018}), \eprint{1804.07365}.

\bibitem[{\citenamefont{Bahamonde
  et~al.}(2018{\natexlab{b}})\citenamefont{Bahamonde, Zubair, and
  Abbas}}]{Bahamonde:2016cul}
\bibinfo{author}{\bibfnamefont{S.}~\bibnamefont{Bahamonde}},
  \bibinfo{author}{\bibfnamefont{M.}~\bibnamefont{Zubair}}, \bibnamefont{and}
  \bibinfo{author}{\bibfnamefont{G.}~\bibnamefont{Abbas}},
  \bibinfo{journal}{Phys. Dark Univ.} \textbf{\bibinfo{volume}{19}},
  \bibinfo{pages}{78} (\bibinfo{year}{2018}{\natexlab{b}}),
  \eprint{1609.08373}.

\bibitem[{\citenamefont{Wright}(2016)}]{Wright:2016ayu}
\bibinfo{author}{\bibfnamefont{M.}~\bibnamefont{Wright}},
  \bibinfo{journal}{Phys. Rev.} \textbf{\bibinfo{volume}{D93}},
  \bibinfo{pages}{103002} (\bibinfo{year}{2016}), \eprint{1602.05764}.

\bibitem[{\citenamefont{Farrugia et~al.}(2020)\citenamefont{Farrugia,
  Levi~Said, and Finch}}]{Farrugia:2020fcu}
\bibinfo{author}{\bibfnamefont{G.}~\bibnamefont{Farrugia}},
  \bibinfo{author}{\bibfnamefont{J.}~\bibnamefont{Levi~Said}},
  \bibnamefont{and} \bibinfo{author}{\bibfnamefont{A.}~\bibnamefont{Finch}},
  \bibinfo{journal}{Universe} \textbf{\bibinfo{volume}{6}}, \bibinfo{pages}{34}
  (\bibinfo{year}{2020}), \eprint{2002.08183}.

\bibitem[{\citenamefont{Capozziello et~al.}(2020)\citenamefont{Capozziello,
  Capriolo, and Caso}}]{Capozziello:2019msc}
\bibinfo{author}{\bibfnamefont{S.}~\bibnamefont{Capozziello}},
  \bibinfo{author}{\bibfnamefont{M.}~\bibnamefont{Capriolo}}, \bibnamefont{and}
  \bibinfo{author}{\bibfnamefont{L.}~\bibnamefont{Caso}},
  \bibinfo{journal}{Eur. Phys. J.} \textbf{\bibinfo{volume}{C80}},
  \bibinfo{pages}{156} (\bibinfo{year}{2020}), \eprint{1912.12469}.

\bibitem[{\citenamefont{Escamilla-Rivera and
  Levi~Said}(2019)}]{Escamilla-Rivera:2019ulu}
\bibinfo{author}{\bibfnamefont{C.}~\bibnamefont{Escamilla-Rivera}}
  \bibnamefont{and} \bibinfo{author}{\bibfnamefont{J.}~\bibnamefont{Levi~Said}}
  (\bibinfo{year}{2019}), \eprint{1909.10328}.

\bibitem[{\citenamefont{Karpathopoulos
  et~al.}(2018)\citenamefont{Karpathopoulos, Basilakos, Leon, Paliathanasis,
  and Tsamparlis}}]{Karpathopoulos:2017arc}
\bibinfo{author}{\bibfnamefont{L.}~\bibnamefont{Karpathopoulos}},
  \bibinfo{author}{\bibfnamefont{S.}~\bibnamefont{Basilakos}},
  \bibinfo{author}{\bibfnamefont{G.}~\bibnamefont{Leon}},
  \bibinfo{author}{\bibfnamefont{A.}~\bibnamefont{Paliathanasis}},
  \bibnamefont{and}
  \bibinfo{author}{\bibfnamefont{M.}~\bibnamefont{Tsamparlis}},
  \bibinfo{journal}{Gen. Rel. Grav.} \textbf{\bibinfo{volume}{50}},
  \bibinfo{pages}{79} (\bibinfo{year}{2018}), \eprint{1709.02197}.

\bibitem[{\citenamefont{Capozziello et~al.}(2015)\citenamefont{Capozziello,
  De~Laurentis, and Myrzakulov}}]{Capozziello:2014bna}
\bibinfo{author}{\bibfnamefont{S.}~\bibnamefont{Capozziello}},
  \bibinfo{author}{\bibfnamefont{M.}~\bibnamefont{De~Laurentis}},
  \bibnamefont{and}
  \bibinfo{author}{\bibfnamefont{R.}~\bibnamefont{Myrzakulov}},
  \bibinfo{journal}{Int. J. Geom. Meth. Mod. Phys.}
  \textbf{\bibinfo{volume}{12}}, \bibinfo{pages}{1550095}
  (\bibinfo{year}{2015}), \eprint{1412.1471}.

\bibitem[{\citenamefont{Pourbagher and Amani}(2019)}]{Pourbagher:2019zhq}
\bibinfo{author}{\bibfnamefont{A.}~\bibnamefont{Pourbagher}} \bibnamefont{and}
  \bibinfo{author}{\bibfnamefont{A.}~\bibnamefont{Amani}},
  \bibinfo{journal}{Astrophys. Space Sci.} \textbf{\bibinfo{volume}{364}},
  \bibinfo{pages}{140} (\bibinfo{year}{2019}), \eprint{1908.11595}.

\bibitem[{\citenamefont{Zubair et~al.}(2018)\citenamefont{Zubair, Waheed,
  Atif~Fayyaz, and Ahmad}}]{Zubair:2018wyy}
\bibinfo{author}{\bibfnamefont{M.}~\bibnamefont{Zubair}},
  \bibinfo{author}{\bibfnamefont{S.}~\bibnamefont{Waheed}},
  \bibinfo{author}{\bibfnamefont{M.}~\bibnamefont{Atif~Fayyaz}},
  \bibnamefont{and} \bibinfo{author}{\bibfnamefont{I.}~\bibnamefont{Ahmad}},
  \bibinfo{journal}{Eur. Phys. J. Plus} \textbf{\bibinfo{volume}{133}},
  \bibinfo{pages}{452} (\bibinfo{year}{2018}), \eprint{1807.07399}.

\bibitem[{\citenamefont{Escamilla-Rivera
  et~al.}(2010)\citenamefont{Escamilla-Rivera, Obregon, and
  Urena-Lopez}}]{EscamillaRivera:2010py}
\bibinfo{author}{\bibfnamefont{C.}~\bibnamefont{Escamilla-Rivera}},
  \bibinfo{author}{\bibfnamefont{O.}~\bibnamefont{Obregon}}, \bibnamefont{and}
  \bibinfo{author}{\bibfnamefont{L.}~\bibnamefont{Urena-Lopez}},
  \bibinfo{journal}{JCAP} \textbf{\bibinfo{volume}{12}}, \bibinfo{pages}{011}
  (\bibinfo{year}{2010}), \eprint{1009.4233}.

\bibitem[{\citenamefont{Shah and Samanta}(2019)}]{Shah:2019mxn}
\bibinfo{author}{\bibfnamefont{P.}~\bibnamefont{Shah}} \bibnamefont{and}
  \bibinfo{author}{\bibfnamefont{G.~C.} \bibnamefont{Samanta}},
  \bibinfo{journal}{Eur. Phys. J. C} \textbf{\bibinfo{volume}{79}},
  \bibinfo{pages}{414} (\bibinfo{year}{2019}), \eprint{1905.09051}.

\bibitem[{\citenamefont{Mirza and Oboudiat}(2017)}]{Mirza:2017vrk}
\bibinfo{author}{\bibfnamefont{B.}~\bibnamefont{Mirza}} \bibnamefont{and}
  \bibinfo{author}{\bibfnamefont{F.}~\bibnamefont{Oboudiat}},
  \bibinfo{journal}{JCAP} \textbf{\bibinfo{volume}{11}}, \bibinfo{pages}{011}
  (\bibinfo{year}{2017}), \eprint{1704.02593}.

\bibitem[{\citenamefont{Odintsov and Oikonomou}(2018)}]{Odintsov:2018uaw}
\bibinfo{author}{\bibfnamefont{S.}~\bibnamefont{Odintsov}} \bibnamefont{and}
  \bibinfo{author}{\bibfnamefont{V.}~\bibnamefont{Oikonomou}},
  \bibinfo{journal}{Phys. Rev. D} \textbf{\bibinfo{volume}{98}},
  \bibinfo{pages}{024013} (\bibinfo{year}{2018}), \eprint{1806.07295}.

\bibitem[{\citenamefont{Said}(2017)}]{Said:2017nti}
\bibinfo{author}{\bibfnamefont{J.~L.} \bibnamefont{Said}},
  \bibinfo{journal}{Eur. Phys. J. C} \textbf{\bibinfo{volume}{77}},
  \bibinfo{pages}{883} (\bibinfo{year}{2017}), \eprint{1712.07592}.

\bibitem[{\citenamefont{Paliathanasis}(2017{\natexlab{b}})}]{Paliathanasis:2017efk}
\bibinfo{author}{\bibfnamefont{A.}~\bibnamefont{Paliathanasis}},
  \bibinfo{journal}{Phys. Rev. D} \textbf{\bibinfo{volume}{95}},
  \bibinfo{pages}{064062} (\bibinfo{year}{2017}{\natexlab{b}}),
  \eprint{1701.04360}.

\bibitem[{\citenamefont{Odintsov and Oikonomou}(2017)}]{Odintsov:2017tbc}
\bibinfo{author}{\bibfnamefont{S.}~\bibnamefont{Odintsov}} \bibnamefont{and}
  \bibinfo{author}{\bibfnamefont{V.}~\bibnamefont{Oikonomou}},
  \bibinfo{journal}{Phys. Rev. D} \textbf{\bibinfo{volume}{96}},
  \bibinfo{pages}{104049} (\bibinfo{year}{2017}), \eprint{1711.02230}.

\end{thebibliography}

\end{document}